\documentclass[]{article}

\usepackage{amsmath}
\usepackage{amssymb}
\usepackage{bbm}
\usepackage{booktabs}
\usepackage{xcolor}
\usepackage{enumitem}
\usepackage{graphicx}
\usepackage{url}
\usepackage{microtype} % necessary?
\usepackage[acronym]{glossaries}
\usepackage{comment}

\usepackage[a4paper, total={6.25in, 9.0in}]{geometry}
\usepackage[skip=10pt plus1pt, indent=0pt]{parskip} % Set paragraph spacing and indent
\usepackage[authoryear]{natbib}
\usepackage[labelfont=bf]{caption}
\usepackage[colorlinks,citecolor=blue]{hyperref}

\newcommand{\one}{\mathbbm{1}}                % indicator function
\newcommand{\R}{\mathbb{R}}                   % real line

\newacronym{NWP}{NWP}{numerical weather prediction}
\newacronym{AIWP}{AIWP}{artificial intelligence weather prediction}
\newacronym{RMSE}{RMSE}{root mean squared error}
\newacronym{CRPS}{CRPS}{continuous ranked probability score}
\newacronym{twCRPS}{twCRPS}{threshold-weighted CRPS}
\newacronym{qwCRPS}{qwCRPS}{quantile-weighted CRPS}
\newacronym{PCRPS}{PCRPS}{potential CRPS}
\newacronym{twPCRPS}{twPCRPS}{threshold-weighted PCRPS}
\newacronym{qwPCRPS}{qwPCRPS}{quantile-weighted PCRPS}

\begin{document}

\title{Towards Fair Comparisons of AI- and Physics-Based Weather Models for Extreme Events via the Weighted Potential CRPS}
\author{Tobias Biegert$^1$, Sam Allen$^1$, Annika Alber$^1$, Sebastian Lerch$^{2,3}$ \\ 
\vspace{0.3cm} \\ 
\small $^1$Institute of Statistics, Karlsruhe Institute of Technology, Karlsruhe, Germany \\
\small $^2$Department of Mathematics and Computer Science, Marburg University, Marburg, Germany \\
\small $^3$Heidelberg Institute for Theoretical Studies, Heidelberg, Germany}
\date{}

\definecolor{darkteal}{rgb}{0,0.35,0.35}
\newcommand{\sam}[1]{\textcolor{darkteal}{\sffamily\small\upshape [Sam: #1]}}
\newcommand{\tobi}[1]{\textcolor{darkteal}{\sffamily\small\upshape [Tobi: #1]}}
\newcommand{\sebastian}[1]{\textcolor{darkteal}{\sffamily\small\upshape [Sebastian: #1]}}

\maketitle

\begin{abstract}
    We study whether deterministic AI weather prediction (AIWP) models issue more informative forecasts for extreme weather events than deterministic numerical weather prediction (NWP) models. The deterministic model output is subjected to statistical post-processing via isotonic distributional regression (IDR), or EasyUQ, before the resulting probabilistic forecasts are assessed using weighted versions of the continuous ranked probability score (CRPS). This extends the recently proposed Potential CRPS (PCRPS) measure to focus on extreme outcomes. Since IDR exhibits optimality properties with respect to weighted versions of the CRPS, the proposed approach inherits desirable properties of the PCRPS, and, in particular, facilitates  fair comparisons between data-driven and physics-based models when forecasting extreme weather events. We apply this evaluation framework to forecasts in the WeatherBench 2 dataset issued by the AIWP models GraphCast, Pangu-Weather, and FuXi, with the ECMWF’s high-resolution NWP model serving as a physics-based reference. The forecast models are compared when predicting mean sea level pressure, temperature, wind speed, and precipitation extremes, defined as exceedances or non-exceedances of thresholds obtained from historical observation data. Across a range of extreme-event thresholds, including thresholds corresponding to record-breaking events, the ranking of the forecast models remains largely unchanged. We find that AIWP models, particularly GraphCast and FuXi, result in the most informative forecasts for extreme weather events across most settings, suggesting that AIWP models have the potential to outperform NWP models when forecasting extremes.
    
\end{abstract}

\section{Introduction}  
\label{sec:introduction}

\Gls{AIWP} models have recently gained considerable attention as an alternative to traditional \gls{NWP} models. While \gls{NWP} systems are governed by the physical laws of the atmosphere, \gls{AIWP} models are based on large-scale machine learning models trained on archives of historical weather data. Although physical constraints are not explicitly enforced, \gls{AIWP} models have been found to perform competitively with state-of-the-art \gls{NWP} models, at a fraction of the computational cost \citep[see e.g.][]{pathak2022fourcastnet,bi2023pangu,lam2023graphcast,chen2023fuxi}. While our primary focus is on deterministic models, recent developments also include probabilistic data-driven weather models capable of generating ensemble forecasts \citep{price2025gencast,zhong2025,lang2026aifs}.

However, a fair comparison of deterministic \gls{AIWP} and \gls{NWP} models is difficult to achieve. Compari employ a range of verification measures, but 
often place particular emphasis on the \gls{RMSE}, which arguably unfairly favours \gls{AIWP} models trained using a squared-error loss; in contrast, \gls{NWP} models are not explicitly trained to minimise a particular loss function. %Indeed, since the MSE is minimised by the conditional mean of the target distribution, \gls{AIWP} models often produce weather fields that are unrealistically smooth; evaluating forecasts using the MSE rewards this rather than penalising it.
Instead, several studies have suggested that a fair comparison can be achieved by first converting the deterministic weather forecasts to probabilistic forecasts, and then evaluating the forecasts using probabilistic verification metrics. \citet{brenowitz2025practical} obtain an ensemble forecast by aggregating deterministic forecasts over time, \citet{loveday2025} similarly aggregate over neighbouring grid points, while \citet{BuelteEtAl2026} and \citet{almeida2025} generate an ensemble by running models from different initial condition perturbations. Alternatively, \citet{gneiting2026pcrps} propose an arguably more objective approach to generate probabilistic forecasts using isotonic distributional regression \citep[IDR;][]{henzi2021isotonic}, or Easy Uncertainty Quantification \citep[EasyUQ;][]{walz2024easyuq}. In all cases, the deterministic \gls{AIWP} and \gls{NWP} model forecasts are compared via the performance of the corresponding probabilistic forecasts, typically using proper scoring rules such as the continuous ranked probability score \citep[CRPS;][]{matheson1976scoring}.\glsunset{CRPS}

While these comparisons tend to focus on overall forecast performance, there has been much interest in the literature regarding how well \gls{AIWP} models can predict extreme weather events. \citet{olivetti2024} compare \gls{AIWP} and \gls{NWP} models using the \gls{RMSE} applied to forecasts made when an extreme event occurs, where an extreme event is defined as the exceedance of a relevant threshold. \citet{zhang2025} adopt a similar approach but focus on record-breaking events, defined as cases where the highest value within a historical archive of data has been exceeded. However, restricting the evaluation to instances where an extreme event occurs is known to favour forecast methods that predict that an extreme event will occur with a higher frequency, resulting in a biased comparison \citep{gneiting2011wcrps,lerch2017dilemma}; this is often referred to as the \emph{forecaster's dilemma}.

To circumvent the forecaster's dilemma when comparing probabilistic forecasts, several weighted scoring rules have been proposed that can focus the evaluation on particular outcomes in a theoretically desirable way \citep{gneiting2011wcrps,diks2011likelihood,holzmann2017focusing,allen2023evaluating}. In this paper, we propose to compare \gls{AIWP} and \gls{NWP} models by first applying EasyUQ to the deterministic forecasts to produce probabilistic forecasts, and then evaluating the forecasts using the threshold-weighted \gls{CRPS} \citep{gneiting2011wcrps}, arguably the most well-known and widely used weighted scoring rule. This allows us to extend the \gls{PCRPS} introduced by \citet{gneiting2026pcrps} to settings where interest lies in extreme outcomes; we refer to the resulting verification measure as the \gls{twPCRPS}. 

A similar approach is suggested by \citet{loveday2025}, who evaluate extreme precipitation forecasts by treating deterministic forecast values at neighbouring grid points as equally weighted ensemble members, and scoring the resulting ensembles using the threshold-weighted CRPS. Our framework differs from theirs primarily in how we construct the predictive distributions. We argue that EasyUQ provides a more objective means to construct probabilistic forecasts than spatial aggregation, since it minimises the in-sample average \gls{CRPS} subject to a weak monotonicity constraint, while also requiring no tuning parameters. 
\citet[Appendix B]{olivetti2024} also compare \gls{AIWP} and \gls{NWP} models using a weighted version of the mean squared error that puts emphasis on extreme events \citep[see][]{Taggart2022}, but, like the standard mean squared error, this scoring function is also minimised by the predictive mean, and such a comparison therefore also arguably favours \gls{AIWP} models trained to target the conditional mean.

Using the \gls{twPCRPS}, we compare four weather models in their ability to predict extreme weather events: we evaluate the physics-based high-resolution deterministic forecast issued by the European Centre for Medium-Range Weather Forecasts' (ECMWF) Integrated Forecasting System (IFS), as well as the \gls{AIWP} models GraphCast \citep{lam2023graphcast}, Pangu-Weather \citep{bi2023pangu}, and FuXi \citep{chen2023fuxi}. The comparison is performed using the WeatherBench 2 benchmark dataset \citep{rasp2024weatherbench}. The models and data are described in detail in the following section. Section~\ref{sec:methods} then introduces the \gls{twPCRPS} more formally, before the results of the comparison are presented in Section~\ref{sec:results}; code to reproduce these results is available at \url{https://github.com/tobiasbiegert/weighted-pcrps}. Finally, Section~\ref{sec:discussion} summarises the main findings and highlights directions for future research.

\section{Data} 

The various forecast models are compared using the WeatherBench 2 benchmark dataset \citep{rasp2024weatherbench}. WeatherBench 2 builds on the original WeatherBench dataset \citep{rasp2020weatherbench}, offering a standardised collection of medium-range (1-14 days) weather forecasts from both \gls{AIWP} and \gls{NWP} models, as well as corresponding verification data. The variables that we consider are 2~m temperature (T2M), measured in Kelvin; 10~m wind speed (WS10), in metres per second;  mean sea level pressure (MSLP), in Pascals; and 24-hour precipitation accumulation (TP24hr), in metres.

We consider forecasts initialised between the 1st January and 16th December 2020. Since FuXi forecasts are not available after this date in WeatherBench 2, this evaluation period ensures that all considered models are evaluated against the same observations, without any missing data. 
The forecasts are initialised at 00 and 12 UTC, leading to $351 \times 2 = 702$ initialisation times. Following \cite{gneiting2026pcrps}, we restrict attention to forecasts issued for lead times of 1, 3, 5, 7, and 10 days. All forecast and verification data used in this study are taken from WeatherBench 2 on a common 1.5$^\circ$ grid, comprised of $240 \times 121 = 29,040$ grid points across the globe. For each variable and each lead time, this leads to $702 \times 29\,040 \approx 20$ million forecast-observation pairs on which to evaluate each model.

\subsection{Verification data}

The \gls{AIWP} and \gls{NWP} models are evaluated by comparing their forecasts to corresponding verification data, often referred to as the ``ground truth''. We perform the comparison using ERA5 global reanalysis fields, which provide a temporally and spatially consistent reconstruction of the atmosphere based on an earlier IFS model and a comprehensive assimilation of historical observations \citep{hersbach2020era5}. ERA5 has become a standard reference in weather and climate research. 

ERA5 reanalysis fields also serve as the primary training dataset for many data-driven forecast models. Since one could argue that this similarly results in an unfair comparison of the AIWP and NWP models, we additionally assessed forecast performance using analysis fields from the ECMWF's operational IFS forecasting system as an alternative verification dataset. These IFS analysis fields are generated using real-time data assimilation, providing the best available estimate of the atmospheric state at the respective initialisation time. We present results using ERA5 data as the ground truth in the main text, and present some results using IFS analysis fields in Appendix~B (see Figure~\ref{fig:tw_pcrps_monthly_records_operational}). The results are broadly consistent for the models and variables included in both comparisons. 

However, both sets of (re)analysis fields only approximate the true state of the atmosphere, given their reliance on model-based analyses and the limited coverage of weather measurements, and therefore exhibit errors themselves. These errors are not accounted for during the evaluation in this study, and we advocate further work on methods to address this, along the lines of \cite{ferro2017measuring} and \cite{bessac2021forecast}. Furthermore, case studies of individual extreme events, potentially with additional verification datasets, could help improve the understanding of potential shortcomings of AIWP and NWP models in specific situations \citep{charlton-perez_etal_2024_ai,PascheEtAl2025,EWBarXiv}.

\subsection{Forecast data}

Four forecast models are compared: the high-resolution deterministic forecast of ECMWF’s Integrated Forecasting System (HRES), and three leading \gls{AIWP} models: GraphCast \citep{lam2023graphcast}, Pangu-Weather \citep{bi2023pangu}, and FuXi \citep{chen2023fuxi}.

The HRES model is the ECMWF’s IFS operational high-resolution deterministic forecast model. At its native resolution, this \gls{NWP} model operates at a horizontal grid spacing of 0.1$^\circ$ (approximately 9~km) with 137 vertical levels, and has been maintained at this resolution since 2016. The model typically undergoes one to two updates per year, which generally lead to slight gains in forecast performance. Forecast cycles start at 00 and 12 UTC and provide predictions up to 10 days into the future. In this study, as in WeatherBench 2, HRES serves as the physical model benchmark against which data-driven approaches are compared.

GraphCast is a Graph Neural Network model that predicts the temporal evolution of the atmosphere using an iterative message passing approach \citep{lam2023graphcast}. The original model is trained on ERA5 reanalysis data from 1979--2019 at 0.25$^\circ$ horizontal resolution, using six upper-air variables on 37 vertical levels together with five surface variables. Forecasts are generated autoregressively by iterating over 6-hourly time steps, which can then be rolled out to longer lead times. Within WeatherBench 2, both the standard ERA5-initialised version and an operational variant initialised with IFS analyses are available. In the following, these will be referred to as GC-ERA5 and GC-IFS, respectively. 

Pangu-Weather is a transformer-based deep learning model for global medium-range weather prediction \citep{bi2023pangu}. It operates on a 0.25$^\circ$ horizontal grid with 13 vertical levels, using five upper-air and four surface variables as inputs. The model was trained on ERA5 reanalysis data for the period 1979--2017, and generates forecasts autoregressively by chaining predictions from models with different lead times (1, 3, 6, and 24 hours). Similarly to GraphCast, both an ERA5-initialised and an IFS-initialised version are provided in WeatherBench 2, denoted by PW-ERA5 and PW-IFS, respectively. Pangu-Weather forecasts are not available for precipitation, and hence only HRES, GraphCast, and FuXi are compared for this variable. 

For the main comparison against ERA5, we use the ERA5-initialised variants GC-ERA5 and PW-ERA5, while the IFS-initialised variants GC-IFS and PW-IFS are used only in the operational comparison in Appendix~B.

FuXi is a cascaded machine learning weather forecasting system based on a U-Transformer architecture \citep{chen2023fuxi}. It generates global forecasts at 6-hourly temporal resolution on a 0.25$^\circ$ grid, and mitigates error accumulation at longer lead times by chaining three separately fine-tuned sub-models optimised for the 0--5\,day, 5--10\,day, and 10--15\,day forecast ranges. The model is trained on 39 years of 6-hourly ERA5 reanalysis data and predicts five upper-air variables on 13 pressure levels, together with five surface variables.

\section{Methods}  
\label{sec:methods}

For each considered model-variable combination, we evaluate the deterministic forecasts using forecast-observation pairs $(x_1, y_1), \dots, (x_n, y_n)$, where $x_1, \dots, x_n \in \R$ are the forecasts issued by the model, and $y_1, \dots, y_n \in \R$ are the corresponding observations. Here, $n = 351 \times 2 = 702$ denotes the size of the evaluation period, and $x_i$ therefore corresponds to the forecast at time $i$, for a specific weather variable, grid point, and lead time. For concision, the weather variable, grid point, and lead time are omitted from the notation. 

\subsection{PCRPS}
\label{sec:methods:subsec:PCRPS}

To allow for a fair comparison of \gls{AIWP} and \gls{NWP} models, \citet{gneiting2026pcrps} propose converting the deterministic forecasts $x_1, \dots, x_n$ to probabilistic forecasts $\hat{F}_1, \dots, \hat{F}_n$ using EasyUQ \citep{walz2024easyuq}, and then evaluating the probabilistic forecasts using the continuous ranked probability score \citep[CRPS;][]{matheson1976scoring}. The \gls{CRPS} for a predictive distribution function $F$ and an observation $y$ can be written as
\begin{align}
	\operatorname{CRPS}(F,y) &= \int_{-\infty}^\infty \left(F(z)-\one_{\{y\leq z\}}\right)^2 dz \\
	&= 2 \int_0^1 \left(\one_{\{y\leq F^{-1}(\alpha)\}}-\alpha\right)\left(F^{-1}(\alpha)-y\right) d\alpha \\
	&= \mathbbm{E}_F|X - y| - \frac{1}{2}\mathbbm{E}_F|X - X'|,
	\label{eq:crps}
\end{align}
where $\one_{\{ \cdot \}}$ denotes the indicator function, $F^{-1}$ is the quantile function of the predictive distribution, and $X$ and $X'$ are independent random variables distributed according to $F$. The first expression shows that the \gls{CRPS} can be expressed as the Brier score \citep{brier1950verification} when evaluating forecast threshold exceedance probabilities, integrated over all thresholds \citep{matheson1976scoring}, whereas the second expression shows that the \gls{CRPS} can also be expressed as the quantile score (or pinball loss) when evaluating quantile forecasts, integrated over all quantiles \citep{laio2007verification}. The final expression shows that the \gls{CRPS} falls into the more general class of kernel scores \citep{gneiting2007strictly}.

EasyUQ applies isotonic distributional regression \citep[IDR;][]{henzi2021isotonic} to the pairs $(x_1, y_1), \dots, (x_n, y_n)$, returning $n$ discrete predictive distributions, $\hat{F}_1, \dots, \hat{F}_n$. Each $\hat F_i$ is a step-function CDF whose support is contained in the observed outcomes and whose jump sizes are determined by the IDR fit. Since these jumps need not be equal, the distributions can be interpreted as weighted ensemble forecasts, with the support points as ensemble members and the jump sizes as their weights. EasyUQ finds the distributions $\hat{F}_1, \dots, \hat{F}_n$ that minimise the average \gls{CRPS} whilst satisfying the constraint that if $x_i \ge x_j$, then $\hat{F}_i$ is larger than $\hat{F}_j$ (in the sense that $\hat{F}_i(z) \le \hat{F}_j(z)$ for all $z \in \R$). The intuition is that a larger deterministic forecast should yield a larger probabilistic forecast. Note that EasyUQ is applied in-sample, directly to the evaluation data; this is done separately for each weather variable, grid point, and lead time. EasyUQ is discussed in detail in Appendix~A.

In this sense, EasyUQ provides an objective means to convert deterministic forecasts to probabilistic forecasts, without requiring additional (possibly subjective) modelling assumptions or hyperparameter choices. Evaluating the EasyUQ forecast distributions using the \gls{CRPS} therefore does not inadvertently favour any of the forecast models, since the forecasts are all optimal with respect to the \gls{CRPS} given the information provided by the deterministic forecast.

Following \citet{gneiting2026pcrps}, we define the potential \gls{CRPS} of a deterministic forecast $x$ as the \gls{CRPS} of the resulting EasyUQ predictive distribution $\hat{F}$,
\begin{equation}
	\operatorname{PCRPS}(x, y) = \operatorname{CRPS}(\hat{F}, y).
	\label{eq:pcrps}
\end{equation}
Since $\hat{F}$ can be interpreted as a weighted ensemble forecast, this can be calculated using any method to compute the \gls{CRPS} for an ensemble forecast \citep{grimit2006continuous,jordan2016facets,zamo2018}. The \gls{AIWP} and \gls{NWP} models can then be compared via their mean $\operatorname{PCRPS}$ over all forecast cases, i.e. 
\begin{equation}
	\overline{\operatorname{PCRPS}} = \frac{1}{n}\sum_{i=1}^{n} \operatorname{PCRPS}(x_i,y_i) = \frac{1}{n}\sum_{i=1}^{n} \operatorname{CRPS}(\hat F_i,y_i).
	\label{eq:mean_pcrps}
\end{equation}

As well as presenting the results for each grid point separately, we additionally present some results aggregated over all grid points. In this case, we display the average \gls{PCRPS} across all grid points, where each grid point is weighted according to the latitude weighting scheme employed by \citet{rasp2024weatherbench}. This is also employed when using the threshold-weighted PCRPS introduced in the following section.

It is additionally convenient to present average scores relative to that of a baseline forecast. These skill scores are generally calculated using the unconditional climatology as a baseline, since this represents a forecast that is calibrated but exhibits no temporal discrimination. In this case, the reference score for the \gls{CRPS} becomes
\begin{equation}
	\overline{\operatorname{PCRPS}}_0 = \frac{1}{n}\sum_{i=1}^n \operatorname{CRPS}(F^0,y_i) = \frac{1}{2n^2}\sum_{i=1}^n\sum_{j=1}^n |y_i-y_j|,
	\label{eq:pcrps_reference}
\end{equation}
where $F^0$ denotes the empirical distribution of $y_1, \dots, y_n$. The \gls{PCRPS} skill score is then defined as
\begin{equation}
	\operatorname{PCRPS-S} = 1 - \frac{\overline{\operatorname{PCRPS}}}{\overline{\operatorname{PCRPS}}_0}.
	\label{eq:pcrps_skill}
\end{equation}
A skill score equal to zero suggests that the forecasting method has the same average \gls{PCRPS} as the climatological baseline forecast, 
while a positive skill score suggests the forecasting method outperforms the baseline. The skill score can thus be interpreted as the relative improvement upon the baseline forecast, with a maximum value of 1 (corresponding to a perfect forecast with $\overline{\operatorname{PCRPS}} = 0$). While skill scores can generally be negative, which would suggest that the forecasting method performs worse than the baseline, this is not possible here since the IDR forecasts are guaranteed to perform no worse than the unconditional climatology $F^0$ in-sample \citep{arnold2024}.

\subsection{Threshold-weighted PCRPS}
\label{sec:methods:subsec:twPCRPS}

To emphasise particular outcomes when evaluating probabilistic forecasts, \citet{gneiting2011wcrps} introduced the \gls{twCRPS}, which incorporates a weight function into the threshold-based representation of the \gls{CRPS},
\begin{align}
	\operatorname{twCRPS}_w(F,y) &= \int_{-\infty}^\infty \left(F(z)-\one_{\{y\leq z\}}\right)^2 w(z) dz \\
	&= \mathbbm{E}|v(X)-v(y)|-\frac{1}{2}\mathbbm{E}|v(X)-v(X')|,
	\label{eq:twcrps}
\end{align}
where $X, X' \sim F$ are independent, $w : \R \to \R_{\ge 0}$ is a non-negative weight function, and $v: \R \rightarrow \R$ is an anti-derivative of $w$, i.e. $v = \int w$, referred to as the chaining function. The second expression demonstrates that the \gls{twCRPS} is equivalent to the \gls{CRPS} applied to a transformed forecast and observation, where the transformation $v$ depends on the weight function $w$ \citep{allen2023evaluating}.

To emphasise extreme weather events, it is common to use an indicator weight function: $w(z)=\one_{\{z > t\}}$ when interest is on values that exceed a threshold $t \in \R$, and $w(z)=\one_{\{z < t\}}$ when interest is on values that fall below the threshold. Corresponding chaining functions are $v(z)=\max\{z,t\}$ and $v(z)=\min\{z,t\}$, respectively. If interest is on the upper tail of the outcome distribution, this yields
\begin{equation}
	\operatorname{twCRPS}_t(F,y) = \int_{t}^\infty \left(F(z)-\one_{\{y\leq z\}}\right)^2 dz,
	\label{eq:twcrps_threshold}
\end{equation}
which demonstrates that the evaluation only concerns the forecast probabilities assigned to values greater than or equal to $t$. An analogous expression exists for the lower tail, with the integration restricted to $(-\infty,t)$.

To compare \gls{AIWP} and \gls{NWP} models with respect to extremes, we extend the PCRPS in Equation~\eqref{eq:pcrps} by defining its threshold-weighted counterpart as
\begin{equation}
	\operatorname{twPCRPS}_t(x, y) = \operatorname{twCRPS}_t(\hat{F}, y),
	\label{eq:twpcrps}
\end{equation}
where a subscript $t$ is used to clarify that this is a function of the threshold $t$. The threshold-weighted PCRPS can similarly be calculated using any method to compute the \gls{twCRPS} for an ensemble forecast \citep{allen2023evaluating}. The forecasts $x_1, \dots, x_n$ can be evaluated using the average $\operatorname{twPCRPS}_t$ over all $n$ forecast cases, 
\begin{equation}
	\overline{\operatorname{twPCRPS}}_t = \frac{1}{n}\sum_{i=1}^{n} \operatorname{twPCRPS}_t(x_i,y_i) = \frac{1}{n}\sum_{i=1}^{n} \operatorname{twCRPS}_t(\hat F_i,y_i).
	\label{eq:mean_twpcrps}
\end{equation}
This average includes cases in which no extreme event occurs. In particular, for upper-tail evaluation, the \gls{twCRPS} penalises any probability that the EasyUQ predictive distribution assigns to values above $t$ when the verifying observation is at or below $t$.

When thresholds vary between forecast cases, we first express forecasts and observations relative to their corresponding thresholds. Writing $t_i$ for the threshold associated with case $i$, we define
\begin{equation}
	(\widetilde x_i,\widetilde y_i)
	=
	\begin{cases}
		(x_i-t_i,y_i-t_i), & \text{for upper-tail evaluation},\\
		(t_i-x_i,t_i-y_i), & \text{for lower-tail evaluation}.
	\end{cases}
	\label{eq:threshold_relative_pairs}
\end{equation}
In both cases, an extreme event corresponds to $\widetilde y_i>0$. EasyUQ is then fitted to the transformed pairs for the chosen tail, yielding predictive distributions for $\widetilde y_i$. We evaluate these distributions against $\widetilde y_i$ using the upper-tail twCRPS at threshold zero, and average the resulting scores over all $n$ forecast cases. The same scoring rule is therefore applied to all transformed cases. We analyse results both when interest is on values above (exceedances) and below (non-exceedances) relevant thresholds; the selected thresholds are discussed in the following subsection. 

As well as optimising the CRPS, the EasyUQ predictive distributions also result in the optimal average \gls{twCRPS} over the test data, subject to the isotonicity constraint \citep[Theorem 2]{henzi2021isotonic}. The transformation described in \eqref{eq:threshold_relative_pairs} ensures that this optimality result holds for varying thresholds. We therefore argue that the threshold-weighted PCRPS similarly provides an objective and fair means to compare \gls{AIWP} and \gls{NWP} models when forecasting extreme weather events. Since the application of EasyUQ essentially statistically post-processes the deterministic forecasts, the resulting forecast distributions can adapt to differences in the observation data, tailoring the evaluation of these forecasts to the choice of observation data. On the other hand, by removing possible biases in the forecasts, the PCRPS only measures the \emph{potential} predictive ability of the deterministic forecasts, and the threshold-weighted PCRPS similarly only measures the potential ability of the deterministic forecasts to predict extreme events. Put differently, the PCRPS can be interpreted as a measure of the forecast discrimination ability or information content, which ignores possible miscalibrations that may arise due to the loss function and data used to train the models; analogously, the threshold-weighted PCRPS can be interpreted as a measure of how informative the deterministic forecast is when predicting extreme outcomes.

A skill score based on the \gls{twCRPS} can similarly be defined, using the empirical distribution $F^0$ of the observations $y_1, \dots, y_n$ as the baseline forecast. For fixed thresholds, the reference threshold-weighted CRPS is
\begin{equation}
	\overline{\operatorname{twPCRPS}}_{0,t} =\frac{1}{n} \sum_{i=1}^n \operatorname{twCRPS}_t(F^0, y_i) = \frac{1}{2n^2} \sum_{i=1}^{n} \sum_{j=1}^{n} |v(y_i) - v(y_j)|.
	\label{eq:twpcrps_reference}
\end{equation}
For varying thresholds, the reference score is computed from the transformed observations $\widetilde y_1, \dots, \widetilde y_n$, using the upper-tail twCRPS with threshold equal to zero. The corresponding skill score is then defined analogously to Equation~\eqref{eq:pcrps_skill}, and is denoted by $\operatorname{twPCRPS-S}_t$.

The threshold-weighted CRPS is not the only weighted scoring rule, and other weighted scoring rules could similarly be employed within this framework. For example, \citet{gneiting2011wcrps} additionally introduced a quantile-weighted CRPS that emphasises different quantiles of the predictive distribution, rather than different thresholds in the outcome space. \citet[Theorem 2]{henzi2021isotonic} demonstrate that EasyUQ is additionally optimal with respect to the quantile-weighted CRPS, facilitating a fair comparison of the post-processed forecast distributions with respect to this weighted scoring rule. In Appendix \ref{app:qwpcrps}, we introduce a quantile-weighted potential CRPS, which is defined analogously to the threshold-weighted PCRPS, and compare the four forecasting models with respect to this score. The quantile-weighted CRPS is generally less useful than the threshold-weighted CRPS since it focuses the evaluation on particular regions of the forecast distribution rather than on particular outcomes, and we therefore restrict attention to the threshold-weighted PCRPS in the main text.

\subsection{Thresholds}
\label{sec:methods:subsec:thresholds}

The threshold-weighted PCRPS can be implemented more generally for any weight function, though we focus here on indicator weight functions that emphasise values above or below a threshold, since this is the simplest and most commonly used approach to define extreme events in practice. To implement the threshold-weighted PCRPS in this case, a suitable threshold (or thresholds) must be chosen that correspond to extreme outcomes. These can then be incorporated into the scoring rule using the weight function $w(z) = \one_{\{z > t\}}$ or $w(z) = \one_{\{z < t\}}$, depending on whether interest is on extremely high or low values.

In the following analysis, we employ thresholds that correspond to high or low quantiles of an archive of ERA5 data from 01-01-1979 to 31-12-2019, which is independent of the data in 2020 that is used for evaluation. The threshold-weighted PCRPS is calculated for thresholds corresponding to the 101 historical quantiles $q \in \{0, 0.01, \dots, 0.99, 1\}$. For temperature and mean sea level pressure, we are interested both in when these thresholds are exceeded, and when the observed value falls below the thresholds. For wind speed and precipitation, we only consider threshold exceedances, since extremely low wind speeds and precipitation accumulations are generally not impactful events (at least not in isolation). These quantiles are calculated separately for each grid point, but over all calendar months. We use these location-specific thresholds to assess extremes relative to the full annual distribution, rather than relative to the time of year. For variables with a pronounced seasonal cycle, such as temperature, the rarity of crossing a given threshold consequently varies throughout the year.

The quantiles $q = 0$ and $q = 1$ correspond to the overall minimum and maximum values of the historical data, therefore defining \emph{overall record} thresholds. We additionally consider \emph{monthly record} thresholds, which are calculated separately for each calendar month. An overall record-breaking event occurs when a value in the evaluation period exceeds or falls below all values in the historical archive, whereas a monthly record-breaking event is defined relative only to historical values from the corresponding calendar month. The monthly record analysis facilitates comparison with \citet{zhang2025}. In contrast to \citet{zhang2025}, we compute the records using historical data up to 2019 rather than 2017, and we evaluate all forecasts using the same verification data (ERA5), rather than evaluating forecasts using the ground truth with which they are initialised (i.e., evaluating AIWP forecasts using ERA5, and NWP forecasts using the IFS analysis). 
The number of monthly records in the evaluation data is presented in Table~\ref{tab:monthly_record_exceedances}, while the records themselves and the frequency of record-breaking events at each grid point are shown in Figures~\ref{fig:q100_thresholds_maps} and~\ref{fig:monthly_record_exceedances_maps} in Appendix~B. 

\begin{table}[t]
	\centering
	\caption{Frequency of monthly record-breaking events for each variable during the evaluation period.}
	\begin{tabular}{@{}lcccccccc@{}}
		\toprule
		& \multicolumn{4}{c}{Number of records} & \multicolumn{4}{c}{Fraction of grid points exhibiting records} \\
		\cmidrule(lr){2-5} \cmidrule(lr){6-9}
		& MSLP & T2M & WS10 & TP24hr & MSLP & T2M & WS10 & TP24hr \\
		\midrule
		Max records & 13603 & 28490 & 7320 & 14417 & 18.07\% & 33.74\% & 19.19\% & 27.16\% \\
		Min records & 7595  & 3500  & --   & --    & 15.26\% & 7.31\%  & --      & --      \\
		\bottomrule
	\end{tabular}
	\label{tab:monthly_record_exceedances}
\end{table}

\section{Results}
\label{sec:results}

\begin{figure}%[t]
	\centering
	\includegraphics[width=\linewidth]{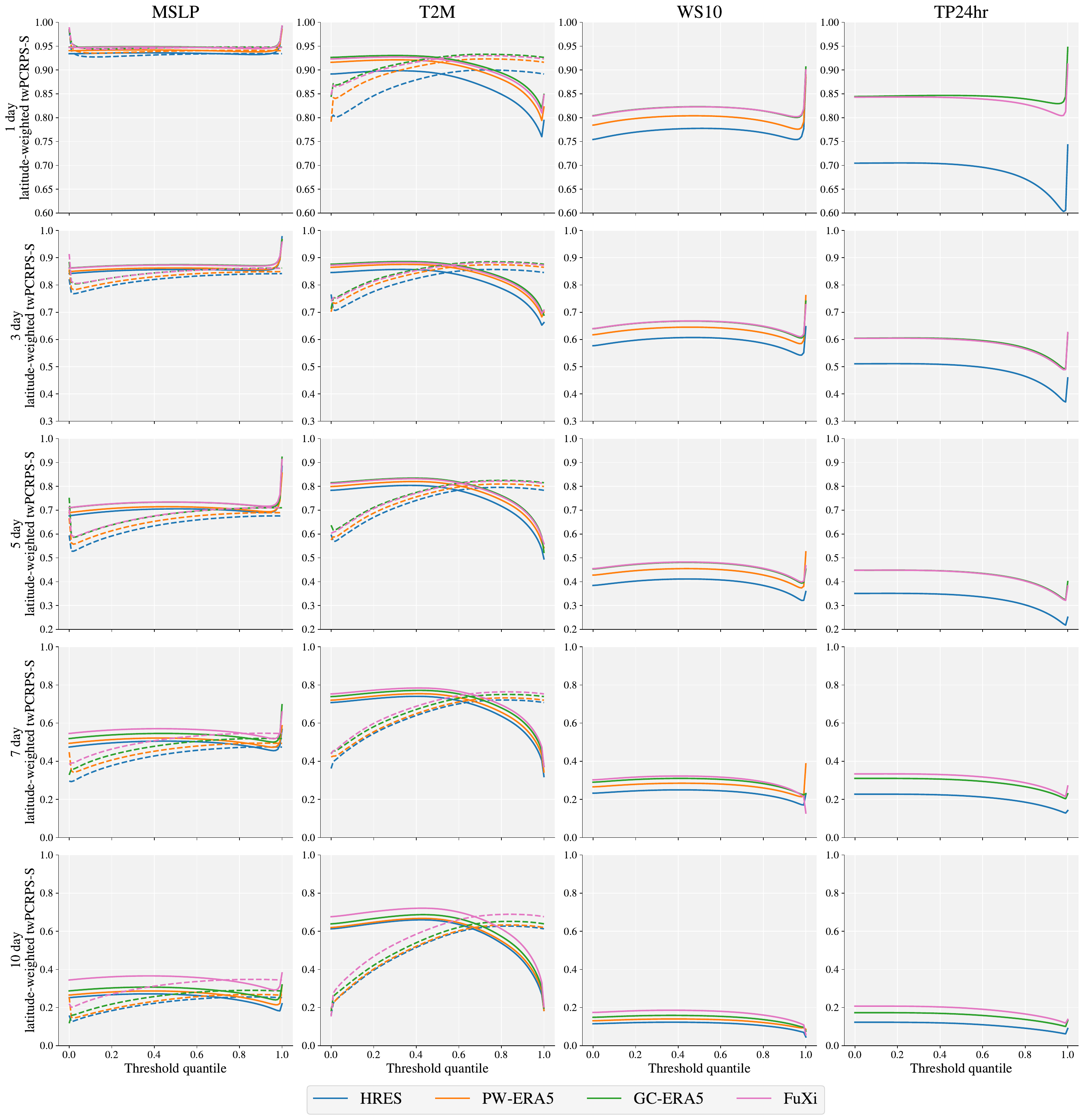}
	\caption{Latitude-weighted $\operatorname{twPCRPS-S}_t$ for the four forecasting models as a function of the quantile of the historical data used to define the threshold in the $\operatorname{twCRPS}_t$. Solid lines correspond to the $\operatorname{twCRPS}_t$ with weight function $w(z) = \one_{\{z > t\}}$, where interest is on threshold exceedances, while dashed lines correspond to the $\operatorname{twCRPS}_t$ with weight function $w(z) = \one_{\{z < t\}}$, where interest is on values not exceeding the threshold. Lower-tail scores are shown only for MSLP and T2M. Columns correspond to weather variables, and rows to lead times. Results are aggregated across all grid points, using ERA5 reanalyses as verification data. For threshold exceedances, a threshold quantile of zero approximately corresponds to the unweighted $\operatorname{PCRPS-S}$; for threshold non-exceedances, the unweighted $\operatorname{PCRPS-S}$ is approximately recovered when the threshold quantile is one. Note that the y-axis scale is the same for all variables but differs across lead times.}
	\label{fig:tw_pcrpss_vs_era5}
\end{figure}

Figure~\ref{fig:tw_pcrpss_vs_era5} displays the $\operatorname{twPCRPS-S}_t$ for the four forecasting models when interest is on values that exceed increasingly high thresholds, as well as values that fall below increasingly low thresholds for MSLP and T2M. Results are shown for all variables and all lead times under consideration. At a threshold of $q = 0$, we are interested in all values that exceed the minimum record in the historical data, in which case the $\operatorname{twPCRPS-S}_t$ is essentially equal to the unweighted $\operatorname{PCRPS-S}$. This facilitates a direct comparison with the results in \citet{gneiting2026pcrps}. The ranking of the different methods tends to be fairly consistent across thresholds, suggesting there is no considerable change in the relative potential skill of AIWP and NWP models when interest is on extreme events. The performance of the four forecast models generally slightly improves relative to the climatological baseline for MSLP, WS10, and TP24hr at very high thresholds, whereas the opposite is true for T2M. As a result, while overall forecast skill is generally highest for MSLP and T2M, and lowest for WS10 and TP24hr, the skill of forecasts for extreme threshold exceedances is generally much higher for MSLP than all other variables.

\begin{figure}%[t]
	\centering
	\includegraphics[width=\linewidth]{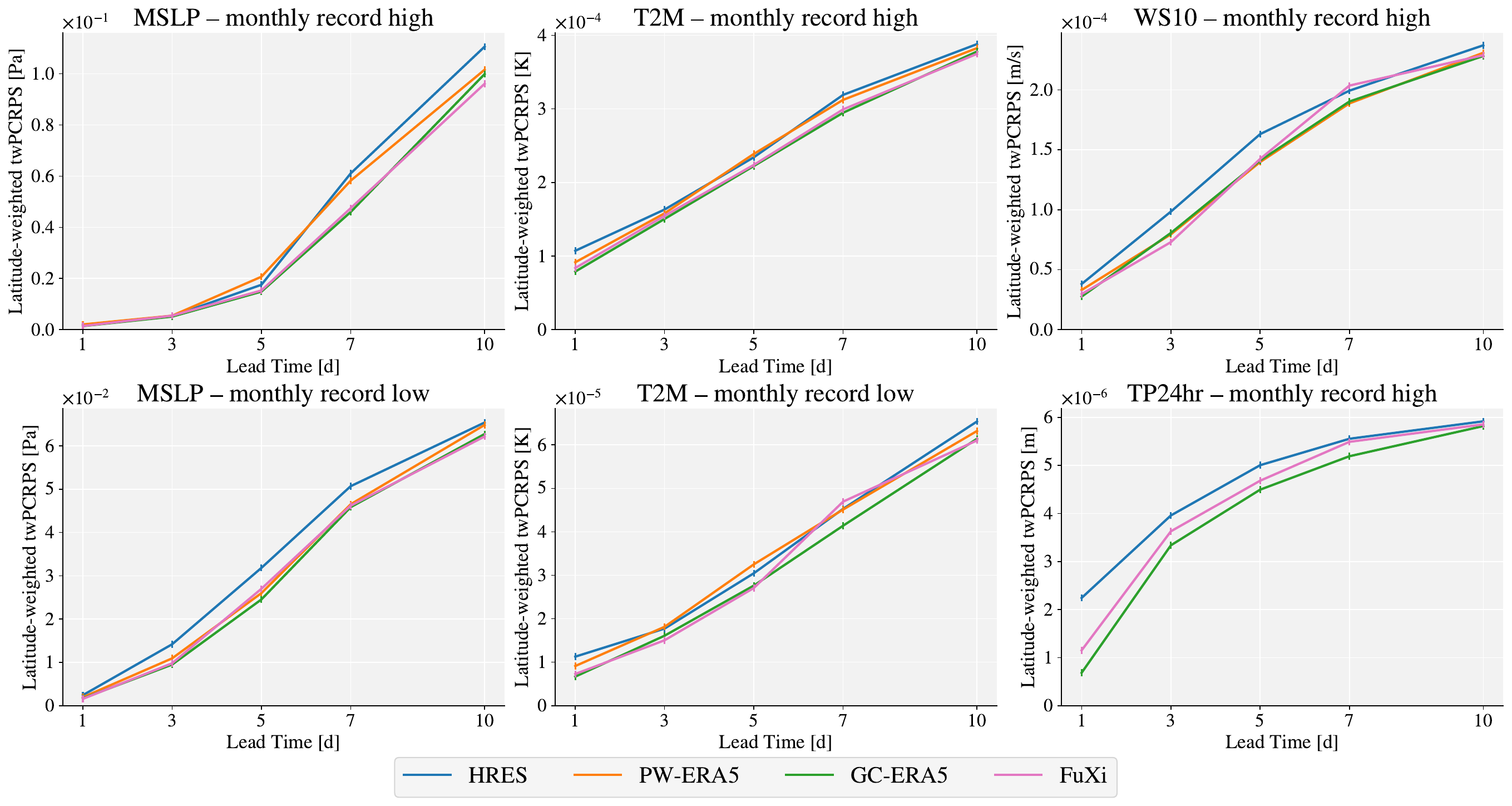}
	\caption{$\overline{\operatorname{twPCRPS}}_t$ for the four forecasting models when interest is on exceedances of record high values and non-exceedances of record low values. Records are calculated for each month separately. Results have been aggregated across all grid points and are displayed as a function of lead time. ERA5 reanalyses are used as verification data.}
	\label{fig:tw_pcrps_monthly_records_vs_era5}
\end{figure}

By analysing results for $q = 1$, we see that the potential accuracy of AIWP models when predicting (overall) record-breaking events is greater than the potential accuracy of NWP models. Corresponding results for monthly records are shown in Figure~\ref{fig:tw_pcrps_monthly_records_vs_era5}. These results similarly favour AIWP models. GraphCast and FuXi generally achieve the best scores, with GraphCast outperforming HRES for every variable, record type and lead time considered. Pangu-Weather is also competitive, particularly for WS10 and monthly record lows of MSLP. For TP24hr, GraphCast consistently achieves the lowest scores, with a particularly pronounced advantage over HRES at short lead times, and smaller differences between the models at longer lead times.
Overall, these results differ from \citet{zhang2025} and are more in line with what \citet{olivetti2024} have found: AIWP models do not exhibit significant drawbacks compared to NWP models when predicting extreme weather events.

\begin{figure}%[t!]
	\centering
	\includegraphics[width=\linewidth]{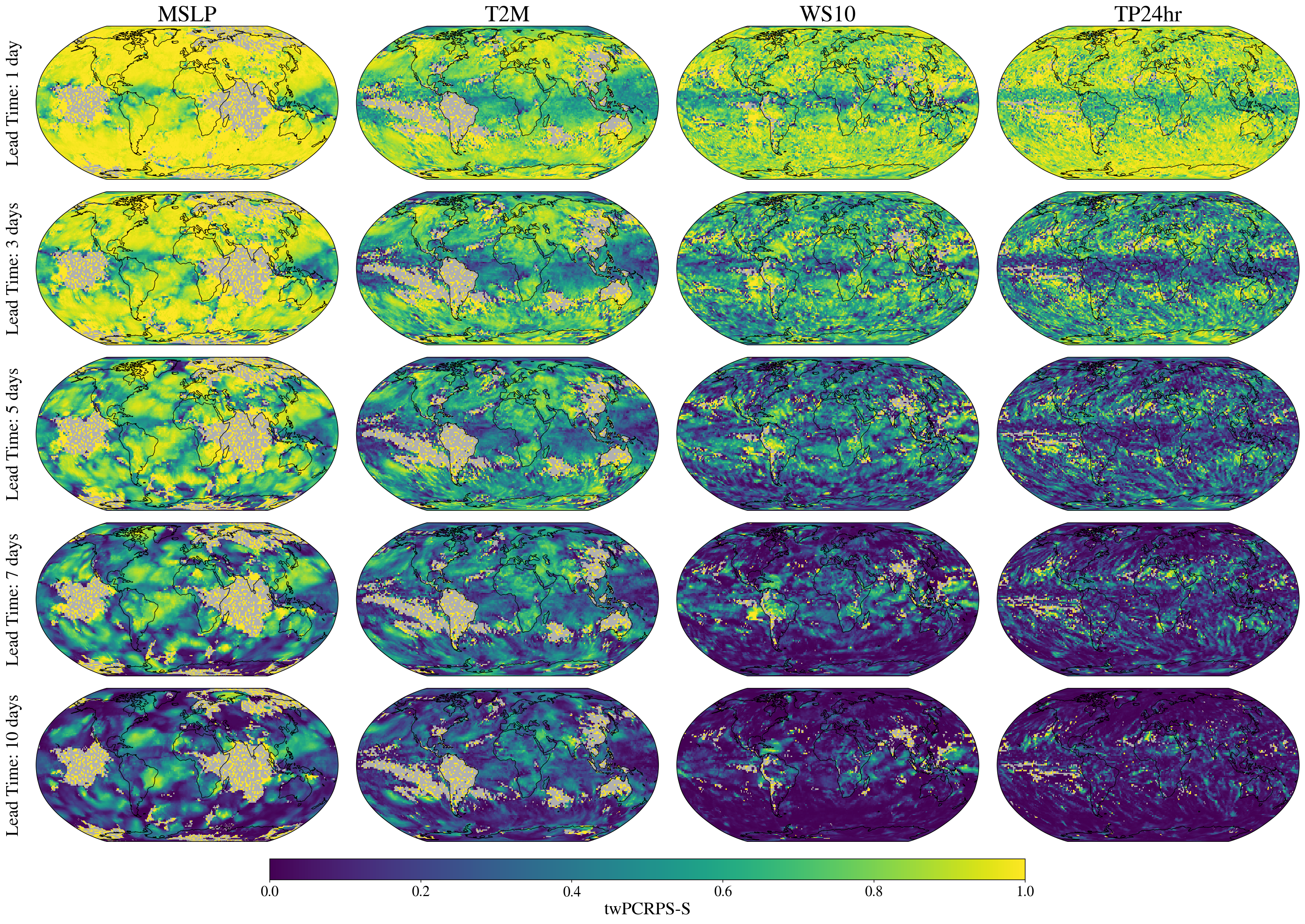}
	\caption{$\operatorname{twPCRPS-S}_t$ of FuXi. Results are shown at each grid point using ERA5 reanalyses as verification data. The twCRPS focuses on exceedances of the historical 99th percentile, computed separately for each grid point. Columns correspond to weather variables and rows correspond to lead times. Skill is measured relative to the $\overline{\operatorname{twPCRPS}}_{0,t}$ baseline. Brighter colours indicate higher potential skill relative to this baseline, while darker colours indicate lower potential skill. Grey grid points indicate locations where the skill score is undefined or numerically unstable because the reference score $\overline{\operatorname{twPCRPS}}_{0,t}$ is zero or very close to zero.}
	\label{fig:maps_fuxi_tw_pcrpss_q099_vs_era5}
\end{figure}

The skill of the FuXi model is shown for each grid point in Figure~\ref{fig:maps_fuxi_tw_pcrpss_q099_vs_era5} when interest is on exceedances of the 99th percentile of the historical data; this threshold is computed separately for each grid point. FuXi exhibits considerable skill when forecasting extreme events at short lead times, though the skill slowly decreases at most grid points as the lead time increases. At short lead times, skill is generally largest in the extratropics and smallest in the tropics, though this pattern varies across weather variables. This is likely due to lower variance in the verification data over the tropics, leaving less scope for AIWP and NWP models to improve upon the climatological baseline.

\begin{figure}%[t!]
	\centering
	\includegraphics[width=\linewidth]{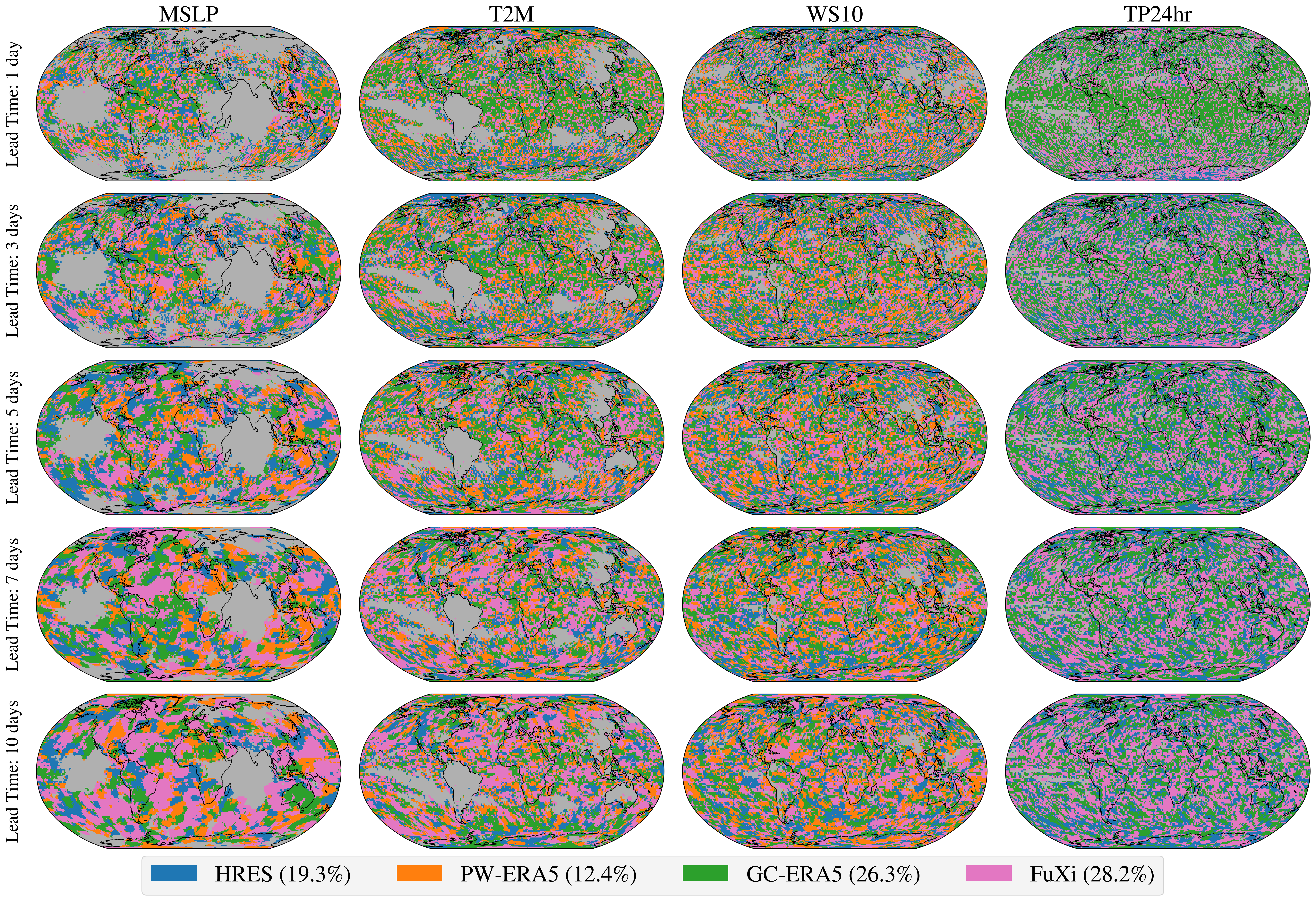}
	\caption{Best-performing model at each grid point according to $\overline{\operatorname{twPCRPS}}_t$ when predicting exceedances of the historical 99th percentile. At each grid point, the colour indicates the model with the lowest $\overline{\operatorname{twPCRPS}}_t$. Columns correspond to weather variables and rows to lead times. The percentages in the legend give the total proportion of cases where each model performs best, aggregated across all grid points, variables, and lead times. Grey grid points indicate ties between two or more models, which primarily occur when multiple models attain a score of zero; these are not assigned to any individual model in the legend percentages. PW-ERA5 is not available for TP24hr and is therefore excluded from that column. ERA5 reanalyses are used as verification data.}
	\label{fig:maps_best_model_tw_pcrps_q099_vs_era5}
\end{figure}

The spatial distribution of forecast skill in Figure~\ref{fig:maps_fuxi_tw_pcrpss_q099_vs_era5} is qualitatively similar for the other forecast models, though FuXi generally exhibits slightly larger skill than the other models, particularly at longer lead times (see Figure~\ref{fig:tw_pcrps_q099_vs_era5} in Appendix~B). The most skilful model at each grid point is displayed in Figure~\ref{fig:maps_best_model_tw_pcrps_q099_vs_era5} when forecasting exceedances of the historical 99th percentile. At short lead times, GraphCast often achieves the lowest $\overline{\operatorname{twPCRPS}}_t$ for these threshold exceedances, particularly for temperature and precipitation. At longer lead times, FuXi performs best over larger regions for MSLP and temperature. For wind speed, the rankings remain spatially mixed, with all four models performing best at many locations. Overall, FuXi and GraphCast are best with similar frequency (28.2\% and 26.3\% of grid points, respectively), and both more often than HRES (19.3\%). Figure~\ref{fig:maps_best_model_tw_pcrps_q099_vs_era5} does not account for whether differences in the $\overline{\operatorname{twPCRPS}}_t$ between models are statistically significant. The significance of these differences can be assessed using block permutation tests, as performed by \citet{gneiting2026pcrps}. Corresponding results are displayed in Figure~\ref{fig:maps_sig_best_model_ai_tw_pcrps_q099_vs_era5} in Appendix~B.

\begin{figure}%[t!]
	\centering
	\includegraphics[width=\linewidth]{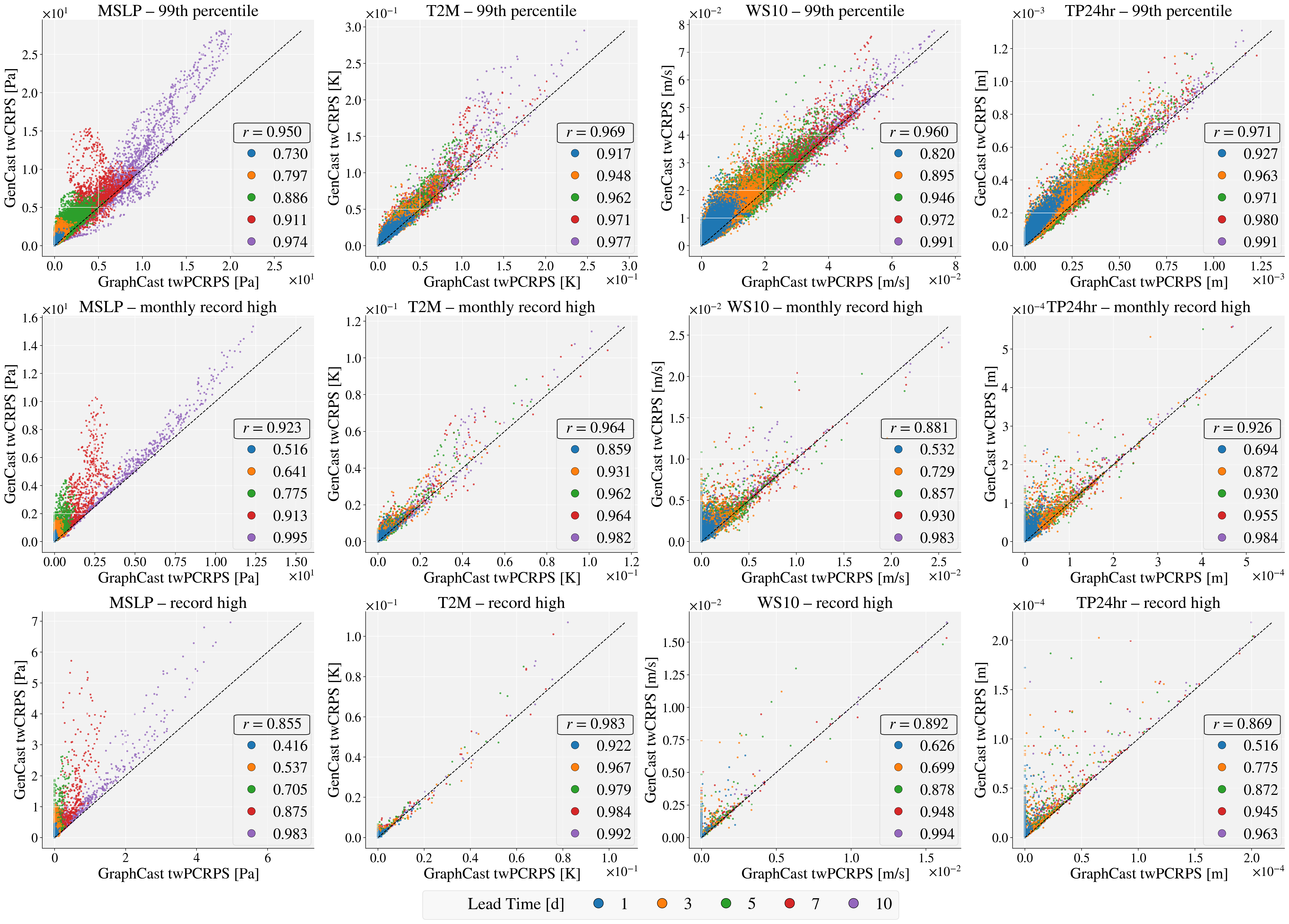}
	\caption{Scatterplots comparing the $\operatorname{twCRPS}_t$ of the GenCast ensemble with the $\operatorname{twPCRPS}_t$ of the deterministic GraphCast forecast. Columns correspond to weather variables and rows to thresholds used in the evaluation: the historical 99th percentile (top), the monthly record threshold (middle), and the overall record threshold (bottom). Points correspond to grid point and lead time combinations, with colours indicating lead time. The dashed diagonal indicates equality between $\operatorname{twPCRPS}_t$ and $\operatorname{twCRPS}_t$, and the inset values report Pearson correlations overall and separately by lead time. ERA5 reanalyses are used as verification data.}
	\label{fig:scatterplot_aiwp_tw_crps_vs_era5}
\end{figure}

As discussed, the $\operatorname{twPCRPS}_t$ quantifies the \emph{potential} accuracy of a deterministic forecast to predict extreme events, essentially evaluating the forecast via the information it contains, rather than the forecast value itself. \citet{gneiting2026pcrps} demonstrate that there is strong correlation between the potential CRPS of a deterministic forecast and the CRPS of a probabilistic forecast constructed from this deterministic backbone; the potential CRPS can therefore be interpreted as a proxy for the CRPS of a probabilistic forecast \citep[see also][]{brenowitz2025practical}. Figure~\ref{fig:scatterplot_aiwp_tw_crps_vs_era5} shows this relationship for GraphCast and its probabilistic counterpart GenCast \citep{price2025gencast}, using upper-tail thresholds given by the historical 99th percentile, the monthly record, and the overall record. Across variables and thresholds, the scatterplots exhibit strong positive correlations, particularly at longer lead times, suggesting that the twPCRPS does indeed provide a good proxy for the relative spatial and lead-time-dependent behaviour of the forecasts, but with a tendency for the twPCRPS to slightly underestimate the twCRPS. The same qualitative pattern is also observed for lower-tail thresholds for mean sea level pressure and 2~m temperature. For comparison, the analogous relationship between HRES and the IFS ensemble is shown in Figure~\ref{fig:scatterplot_nwp_tw_crps_vs_era5} in Appendix~B.

To assess how much of this potential can be realised by applying EasyUQ out of sample, we also report a holdout experiment for PW-ERA5 and HRES in Appendix~B. We apply EasyUQ fitted to 2018--2019 data to the 2020 forecasts and compare the resulting out-of-sample twCRPS with the in-sample twPCRPS on the same cases. The model ranking is largely preserved at the 1st- and 99th-percentile thresholds (Figure~\ref{fig:tw_pcrps_q099_vs_era5_oos}). The relative gaps between the scores are similar for the two models across most thresholds, and generally widen for more extreme thresholds, particularly at short lead times (Figure~\ref{fig:tw_pcrpss_vs_era5_oos}).

\section{Discussion}  
\label{sec:discussion} 

In this work, we extend the potential CRPS measure introduced by \citet{gneiting2026pcrps} to allow for a fair comparison of \gls{AIWP} and \gls{NWP} models when interest is on extreme weather events. Like \citet{gneiting2026pcrps}, we propose using EasyUQ to obtain probabilistic forecasts from the deterministic model output. We then evaluate the EasyUQ predictive distributions using the threshold-weighted CRPS \citep{gneiting2011wcrps}, which allows us to focus the evaluation on extreme outcomes. Since EasyUQ results in optimal forecasts with respect to the threshold-weighted CRPS  when the same scoring rule is applied to all cases \citep[Theorem 2]{henzi2021isotonic}, the resulting threshold-weighted potential CRPS shares the same theoretical properties as the potential CRPS. We therefore argue that the threshold-weighted PCRPS enables a fair benchmarking of deterministic \gls{AIWP} and \gls{NWP} model output with explicit focus on extremes.

We used this framework to compare how well leading \gls{AIWP} and \gls{NWP} models can predict extreme weather events. This comparison was performed using the WeatherBench 2 dataset \citep{rasp2024weatherbench}, across several weather variables over the entire globe. Across most settings, \gls{AIWP} models displayed superior potential predictive ability compared to the physics-based model. Among the \gls{AIWP} models, FuXi and GraphCast generally yielded the best performance. The results for the threshold-weighted PCRPS were generally consistent with those for the PCRPS in \citet{gneiting2026pcrps}, and were largely insensitive to the chosen threshold.

These results suggest that AIWP models have the potential to provide more accurate forecasts for extreme and record-breaking events than NWP models. This finding is in line with conclusions drawn by \citet{olivetti2024}, but in contrast to those drawn by \citet{zhang2025}. Our results are not directly comparable to those in the main analysis of \citet{zhang2025}, since we evaluate all models using the same verification data, whereas \cite{zhang2025} evaluate the HRES model using IFS analyses as ground truth, and AIWP models using ERA5 reanalyses as ground truth. \cite{zhang2025} also report an operational comparison using common IFS verification data, in which HRES still outperforms the AIWP models.
However, a likely explanation for the discrepancy in our conclusions is that the twPCRPS measures the information content of the forecast when predicting extreme events, rather than directly assessing the accuracy of the deterministic forecasts themselves. Since squared-error loss encourages predictions of the conditional mean \citep[see e.g.][]{gneiting2011making}, AIWP models trained using such losses may exhibit reduced variability and therefore under-represent extreme values. Hence, due to the forecaster's dilemma, these models may appear to perform poorly when evaluation is conditioned on an extreme event having occurred. However, there could still be information in the deterministic forecasts, in the sense that larger predictions tend to be associated with larger outcomes, even when interest is on extremes. Since this is what is assessed using the twPCRPS, this would explain why AIWP models are found to outperform NWP models when assessed using this metric.

In parallel work \citep{biegert2026}, we test this explanation by comparing operational variants of Pangu-Weather, GraphCast, and Aurora with HRES for monthly record-breaking heat, cold, and wind events in 2022, with all four models initialised from, and verified against, the same analysis fields. Conditioning the evaluation on observed record-breaking events, as in \citet{zhang2025}, favours HRES, whereas under the twPCRPS, HRES performs best at short lead times but no longer consistently outperforms the AI models. Under the threshold-weighted squared error of \citet{Taggart2022}, HRES performs worst at longer lead times. A contingency table analysis shows that the AI models forecast record-breaking events less often than HRES, so misses dominate their threshold-weighted errors and false alarms dominate those of HRES, while score decompositions attribute the differences in rankings mainly to miscalibration rather than discrimination ability.

The threshold-weighted PCRPS can also be interpreted as a proxy for the twCRPS of a probabilistic forecast based on the same model architecture. These results therefore additionally suggest that probabilistic forecasts based on AIWP models should be able to outperform existing ensemble prediction systems when predicting extreme weather. Further work could verify this by comparing existing physics-based ensemble prediction systems with probabilistic AIWP model output \citep[e.g.][]{NeuralGCM,price2025gencast,zhong2025,bonev2025fourcastnet,AletEtAl2025,lang2026aifs,NvidiaAtlas_preprint} when forecasting extreme events. This could include both univariate and multivariate weighted scoring rules, as well as checks for forecast tail calibration \citep{allen2023weighted}.

While the twPCRPS only evaluates the potential predictive performance of a forecasting system, it still provides useful information for operational forecasters using deterministic forecasts as numerical guidance. In particular, the twPCRPS quantifies the information content or discriminative ability of the forecasts, ignoring systematic miscalibration errors. This is similar in essence to the anomaly correlation coefficient (ACC), which is widely used for the evaluation of deterministic forecasts, and which also provides a measure of discriminative ability that is insensitive to systematic forecast biases. While the ACC is inherently connected to the squared error, the twPCRPS circumvents this by first converting the deterministic forecasts to probabilistic forecasts, and is thus less sensitive to the loss function used to train the AIWP models. A forecaster using a deterministic temperature forecast as numerical guidance may see that the issued forecast falls below a warning threshold, suggesting that the observed temperature will not be extreme. However, if they are aware that similar forecasts have often been followed by extreme temperature observations in the past, due to systematic biases in the forecasts, then they can recognise that there is a non-negligible probability that an extreme event will occur after receiving this forecast. That is, the deterministic forecast provides information regarding whether an extreme event will occur, even if a systematic bias means the occurrence or severity of these events are inaccurately predicted. This information content of the deterministic forecasts is what the twPCRPS evaluates.

Our results concern the predictive information extracted from individual forecast values on a common $1.5^\circ$ grid. Regridding may remove information, with potentially different effects across models with different native resolutions. Neighbouring forecast values may also provide additional information that the local EasyUQ fits do not use, particularly for precipitation. The results therefore do not quantify all predictive information contained in the original model output. Future work could assess sensitivity to the evaluation resolution, and investigate the use of IDR with neighbouring forecast values as additional covariates.

A further limitation is that our evaluation assesses univariate predictive distributions and therefore does not measure spatial coherence, temporal consistency, or dependencies between different weather variables. However, high-impact weather events often occur as the result of multiple confounding weather hazards, referred to as compound weather events \citep{zscheischler2020typology}. There is currently no similar objective approach to construct optimal multivariate probabilistic forecasts from deterministic model output. Nevertheless, \citet{loveday2025} and \citet{brenowitz2025practical} propose generating probabilistic forecasts from deterministic weather models using spatial and temporal aggregation. While this approach depends on the (arguably subjective) selection of additional hyperparameters, it could also be used to compare the multivariate performance of deterministic AIWP and NWP models, possibly with a focus on extreme weather events.

\section*{Acknowledgements}

Tobias Biegert gratefully acknowledges support by the German Weather Service (Deutscher Wetterdienst) through the SPARC-ML project within the extramural research programme, funding reference number 4823EMF01. Sebastian Lerch and Sam Allen gratefully acknowledge support by the Vector Stiftung through the Young Investigator Group ``Artificial Intelligence for Probabilistic Weather Forecasting''. We thank Amy McGovern and three anonymous referees for their constructive comments. We also thank Tilmann Gneiting, Sebastian Engelke, and Zhongwei Zhang for fruitful discussions.

\begingroup
\interlinepenalty=10000
\bibliographystyle{apalike}
\bibliography{references}
\endgroup  

\appendix

\appendix
\section{Easy Uncertainty Quantification (EasyUQ)}
\label{app:easyuq}

The \gls{PCRPS} and \gls{twPCRPS} correspond respectively to the \gls{CRPS} and \gls{twCRPS} applied to predictive distributions obtained using Easy Uncertainty Quantification \citep[EasyUQ;][]{walz2024easyuq}, a special case of Isotonic Distributional Regression \citep[IDR;][]{henzi2021isotonic}. IDR is a nonparametric distributional regression method that estimates the conditional distribution of a univariate outcome variable under the assumption that there is an isotonic relationship between the covariates and the outcome; that is, larger covariate values correspond to stochastically larger outcomes. EasyUQ corresponds to the case where the covariate is a single deterministic forecast, in which case it provides a simple means to convert deterministic forecasts to probabilistic forecasts, without requiring any additional assumptions or hyperparameter choices.

Given forecast-observation pairs $(x_1, y_1), \dots, (x_n, y_n)$ with deterministic forecasts $x_i \in \R$ and corresponding outcomes $y_i \in \R$, EasyUQ seeks distributions $\hat{F}_1, \dots, \hat{F}_n$ that are stochastically ordered according to the ordering of the forecasts $x_1, \dots, x_n$. That is, 
\[
x_i \le x_j \implies \hat{F}_i \le_{\text{st}} \hat{F}_j \quad \text{for all $i, j \in \{1, \dots, n\},$}
\]
where $\hat{F}_i \le_{\text{st}} \hat{F}_j$ means that $\hat{F}_i(z) \ge \hat{F}_j(z)$ for all $z \in \R$, or equivalently, $q_\alpha(\hat{F}_i) \le q_\alpha(\hat{F}_j)$ for all $\alpha \in (0, 1)$, where $q_\alpha(F)$ denotes the $\alpha$-quantile of $F$.

The EasyUQ estimator is then defined as the unique solution to a constrained optimisation problem that minimises the mean \gls{CRPS} across all admissible distributions \citep[Theorem~1]{henzi2021isotonic}. That is, it finds the distributions $\hat{F}_1, \dots, \hat{F}_n$ that minimise
\[
\frac{1}{n} \sum_{i=1}^n \operatorname{CRPS}(\hat{F}_i, y_i)
\]
subject to the isotonicity constraint $x_i \le x_j \implies \hat{F}_i \le_{\text{st}} \hat{F}_j$. The resulting predictive distributions are discrete, supported on the previously observed values $y_1, \dots, y_n$, and can be efficiently computed using pool-adjacent-violators (PAV) type algorithms \citep{henzi2021isotonic}. \citet[Theorem 2]{henzi2021isotonic} demonstrate that the EasyUQ predictive distributions additionally minimise the average threshold- and quantile-weighted CRPS, for any choices of the weight functions, provided that the same scoring rule is applied to all cases.

In the framework introduced by \cite{gneiting2026pcrps}, the EasyUQ predictive distributions are constructed by applying EasyUQ to the forecast-observation pairs in the evaluation period. That is, the post-processing procedure is performed in-sample on the test data. This ensures that the predictive distributions have the optimal performance on the test data, and means the evaluation is not dependent on external datasets.

\subsection{Time-varying thresholds}
\label{app:easyuq:subsec:varying_thresholds}

The transformation in Equation~\eqref{eq:threshold_relative_pairs} changes the covariates as well as the outcomes used to fit EasyUQ. The stochastic ordering assumption now associates larger forecast departures from the thresholds with stochastically larger outcome departures, in the direction of the relevant extreme.

When the threshold is constant across all cases within a fit, translation, together with sign reversal for lower-tail evaluation, yields the same score as fitting the original pairs and evaluating the chosen tail at that threshold. A single original EasyUQ fit can therefore be reused for all fixed thresholds. For varying thresholds, the case-specific transformation generally changes the regression problem and can change the ordering of the forecasts. Evaluating the original fit at varying thresholds does not generally retain the optimality guarantee. We instead fit EasyUQ to the transformed forecasts and observations. The resulting predictive distributions minimise the average upper-tail twCRPS at threshold zero, subject to the stochastic ordering constraint on the transformed forecasts.

\subsection{Out-of-sample prediction}
\label{app:easyuq:subsec:out_of_sample}

Out-of-sample predictions can be obtained by fitting EasyUQ to training forecast-observation pairs and applying the fitted mapping to the deterministic forecasts in a separate set of evaluation pairs. If an evaluation forecast matches a training forecast value, the mapping returns the corresponding fitted CDF. For intermediate values, it linearly interpolates the fitted CDFs at the nearest training forecast values below and above the evaluation forecast. Outside the range of training forecasts, it returns the fitted CDF at the nearest endpoint \citep[Section~2.5]{henzi2021isotonic}. For time-varying thresholds, the mapping is fitted to the transformed training pairs defined in Equation~\eqref{eq:threshold_relative_pairs} and applied to the transformed evaluation forecasts.

For a fixed threshold, the mean \gls{twPCRPS} is a lower bound on the mean out-of-sample \gls{twCRPS} computed on the same evaluation pairs. The prediction rule preserves stochastic ordering, so the bound follows from the optimality of the in-sample fit \citep[Theorem~2]{henzi2021isotonic}. For time-varying thresholds, the same bound holds on the transformed scale at threshold zero.

A limitation of EasyUQ for predicting extremes out of sample is that its predictive distributions are supported only on the training observations. When these observations are part of the archive used to define the historical records, the predictive distributions therefore assign zero probability to breaking a record. For both overall and monthly records, the out-of-sample predictive CDF $\widetilde F_i(z)$ on the record-relative scale of Equation~\eqref{eq:threshold_relative_pairs} then equals one for $z\geq 0$. Applying Equation~\eqref{eq:twcrps_threshold} at threshold zero gives the out-of-sample score

\begin{equation}
	\operatorname{twCRPS}_0(\widetilde F_i,\widetilde y_i)
	= \int_0^\infty \left(1-\one_{\{\widetilde y_i\leq z\}}\right)^2\,dz
	= \int_0^\infty \one_{\{z<\widetilde y_i\}}\,dz
	= \max\{\widetilde y_i,0\}.
	\label{eq:twcrps_oos_records}
\end{equation}
The out-of-sample score thus depends only on the observed departure from the record and is identical across models evaluated on the same cases using the same record thresholds.

The discrete structure of EasyUQ also induces systematic departures from probabilistic calibration when predicting continuous outcomes out of sample. \citet{biegert2026calibration} document spikes at simple fractions in the distribution of probability integral transform (PIT) values and explain how they arise from the pooling of observations in isotonic regression.

\subsection{Computational details}
\label{app:easyuq:subsec:computation}

We implement EasyUQ using the accelerated PAV algorithm of \citet{henzi2022accelerating}. The computational cost of fitting EasyUQ to $n$ forecast-observation pairs has an $O(n^2)$ upper bound \citep[Section 3.4]{walz2024easyuq}. Computing the mean PCRPS from the fitted distributions requires up to $O(n^2)$ additional operations \citep{gneiting2026pcrps}. Fits are computed separately for each model, variable, grid point, and lead time. Spatial chunks and different variables are processed in parallel. The original fitted distributions are reused when calculating the PCRPS, qwPCRPS, and twPCRPS for all 101 fixed thresholds. Two additional fits are computed for the monthly record minimum and maximum thresholds using the corresponding transformed pairs.

Score computations were performed with two Intel Xeon E5-2680 v3 processors, using the Apache Beam DirectRunner with 40 worker processes. For the main ERA5 comparison with $n=702$ cases per fit, fitting EasyUQ and constructing the predictive distributions from the original pairs took an average of $0.232$~s. The additional fits for the monthly record minimum and maximum thresholds took $0.232$ and $0.231$~s, respectively. Evaluating these cases required an additional $0.024$~s for the PCRPS, $0.164$~s for the qwPCRPS across 99 quantile levels, and $0.224$~s for the twPCRPS across all 103 threshold settings. These measurements exclude data loading and output writing.

The four forecast-model comparisons against ERA5 took a total of $18.03$~h of wall-clock time. The three operational comparisons against IFS analyses took $10.61$~h. The nine pairwise block permutation-test jobs took a further $8.98$~h. These wall-clock times include data loading and output writing. Observed peak memory usage was $230.9$~GB during the scoring run and $44.6$~GB during the permutation tests.
 
\section{Additional results}
\label{app:add_results}

\begin{figure}
	\centering
	\includegraphics[width=\linewidth]{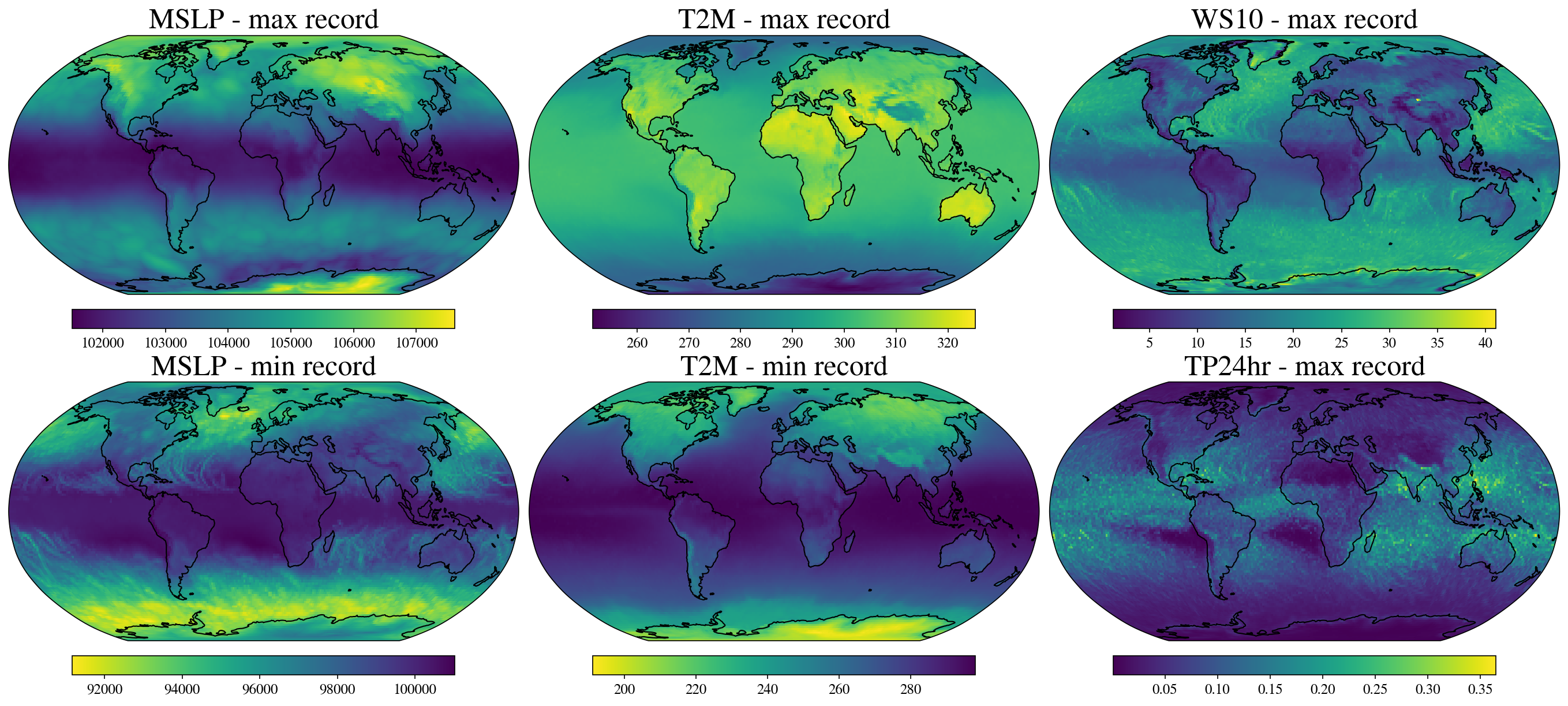}
	\caption{Historical record thresholds at each grid point in the ERA5 reanalysis data from 1979 to 2019. Both maximum and minimum values are shown for mean sea level pressure and 2~m temperature, while only maximum values are shown for 10~m wind speed and 24-hour precipitation accumulation. The colour scale differs between panels, with yellow values always denoting more extreme values.}
	\label{fig:q100_thresholds_maps}
\end{figure}

Figure~\ref{fig:q100_thresholds_maps} displays the (overall) record values of each variable at every grid point in the historical archive of ERA5 data. Results are shown for both the minimum and maximum values of MSLP and T2M, allowing us to analyse both extremely high and extremely low events, while only the maximum values of WS10 and TP24hr are displayed, since the low extremes (i.e. essentially wind speed or precipitation equal to zero) are not of interest here. Unsurprisingly, temperature records are generally highest in the tropics, where MSLP record highs are lowest and MSLP record lows are highest. Precipitation records are similarly largest in the tropics, though this pattern is more intermittent, while wind speed records are generally larger over ocean than over land.

\begin{figure}
	\centering
	\includegraphics[width=\linewidth]{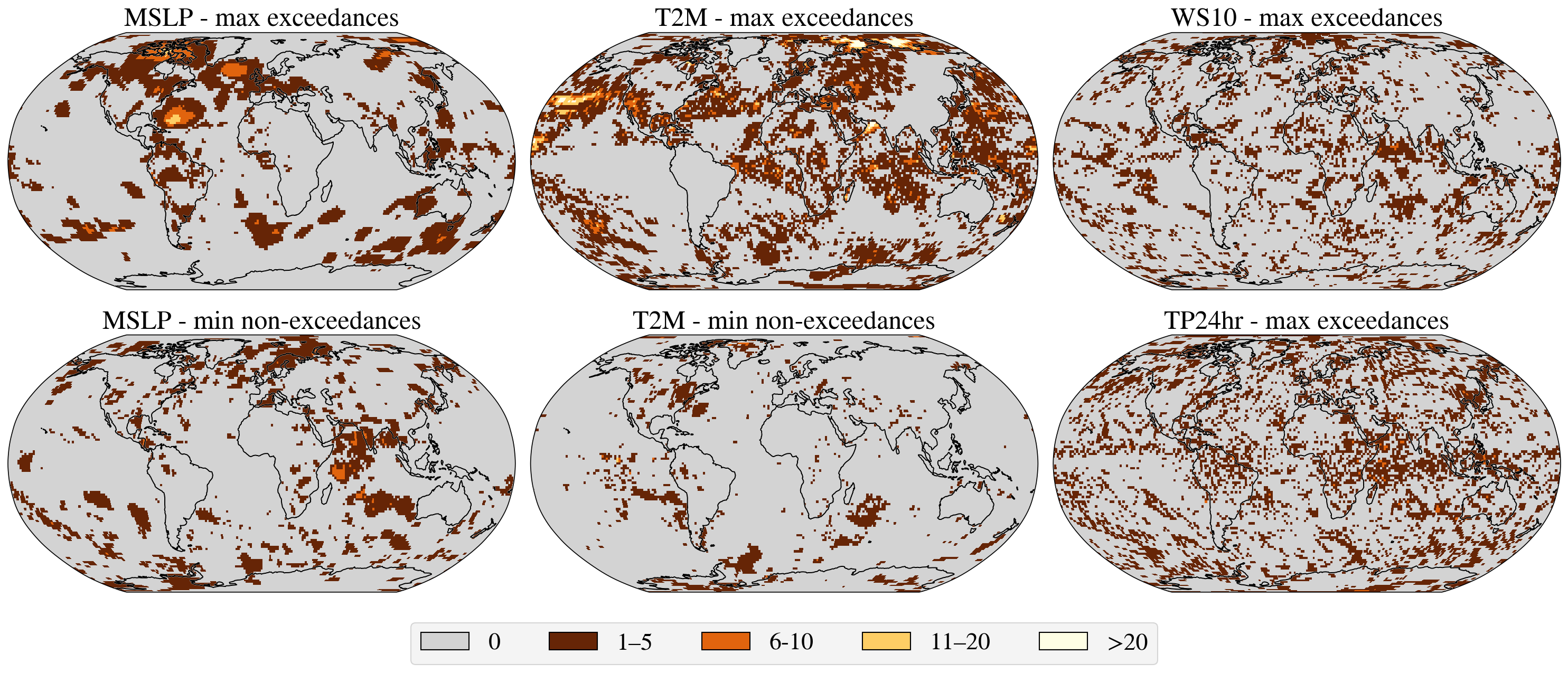}
	\caption{The total number of exceedances of monthly record high values and non-exceedances of monthly record low values at each grid point during the evaluation period.}
	\label{fig:monthly_record_exceedances_maps}
\end{figure}

Figure~\ref{fig:monthly_record_exceedances_maps} shows the number of times that a monthly record is exceeded (or not exceeded, in the case of extreme low values) at each grid point. Perhaps unsurprisingly, record high temperatures were exceeded most frequently, more than 70 times at some grid points, and record low temperatures were broken least frequently; this pattern holds for the majority of grid points. MSLP records were broken at clusters of locations, such as in the North Atlantic Ocean and Indian Ocean, whereas WS10 and TP24hr records were broken less often and more sporadically.

\begin{figure}[t!]
	\centering
	\includegraphics[width=\linewidth]{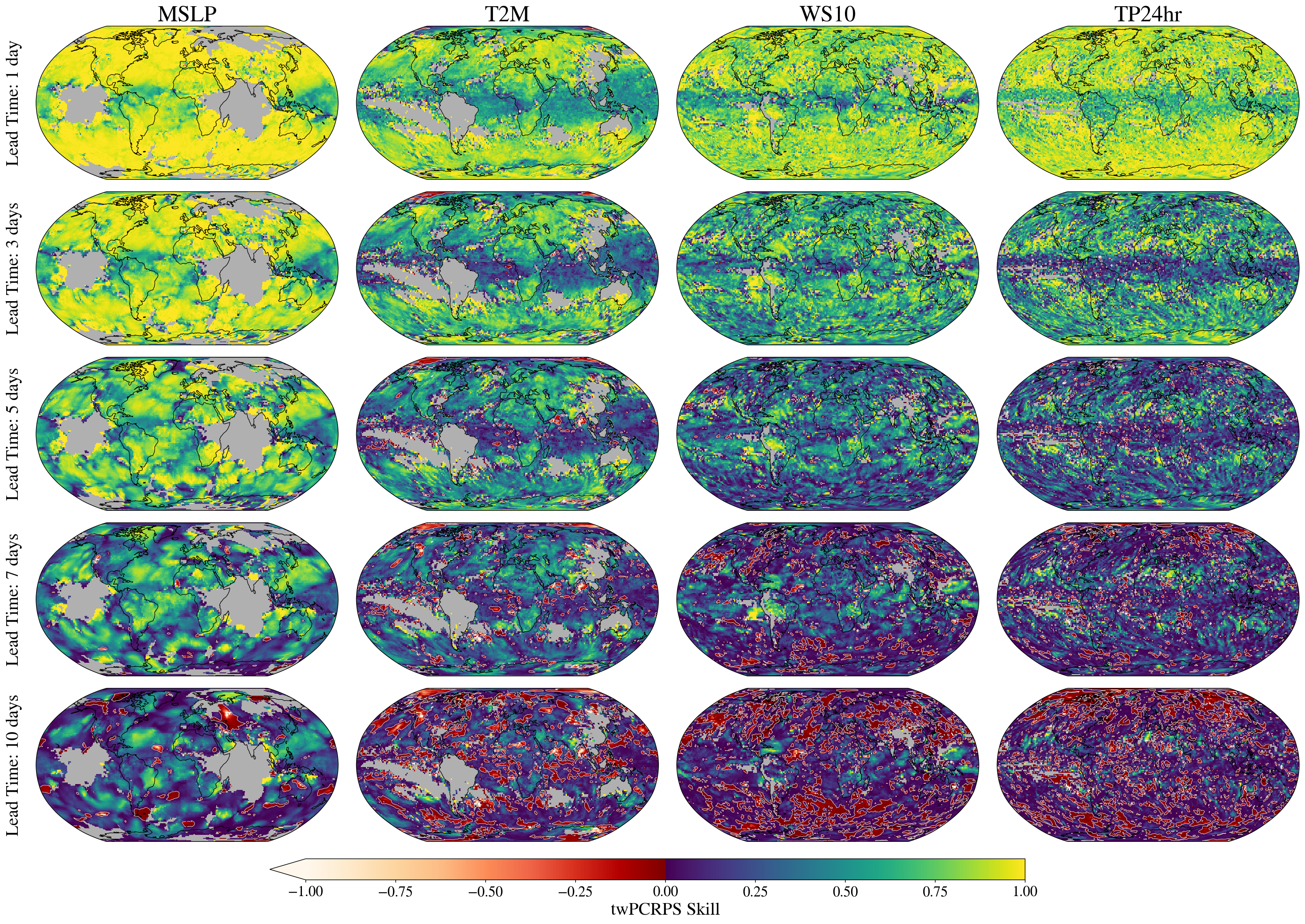}
	\caption{$\operatorname{twPCRPS}_t$ skill of FuXi relative to the seasonally varying ERA5 climatology forecast from WeatherBench 2. Results are shown at each grid point using ERA5 reanalyses as verification data. The threshold-weighted score focuses on exceedances of the historical 99th percentile, computed separately for each grid point. Columns correspond to weather variables and rows correspond to lead times. Positive values indicate improvement over the climatological baseline, while negative values indicate worse performance. Grey grid points indicate locations where the skill score is undefined or numerically unstable because the reference score is zero or very close to zero.}
	\label{fig:maps_fuxi_tw_pcrps_skill_q099_vs_era5}
\end{figure}

While Figure~\ref{fig:maps_fuxi_tw_pcrpss_q099_vs_era5} displays the $\operatorname{twPCRPS-S}_t$ for FuXi at each grid point, one could argue that the unconditional climatology is a relatively weak baseline forecast. Instead, it is common to additionally evaluate forecast performance using a seasonal climatology as a baseline; in WeatherBench 2, this is available via the ERA5 climatology forecast. Figure~\ref{fig:maps_fuxi_tw_pcrps_skill_q099_vs_era5} therefore displays the $\operatorname{twPCRPS}_t$ skill of FuXi forecasts relative to this seasonal ERA5 climatology forecast when interest is on exceedances of the historical 99th percentile; this is calculated and displayed separately for each grid point. The results are qualitatively similar to those in Figure~\ref{fig:maps_fuxi_tw_pcrpss_q099_vs_era5}, with lower improvements over the tropics for all variables at short lead times. Forecast skill again decreases with lead time, though, unlike the $\operatorname{twPCRPS-S}_t$, the skill relative to the seasonal ERA5 climatology is not constrained to be positive, and we observe negative skill scores at many grid points at a lead time of 10 days for all weather variables. This suggests that FuXi forecasts do not improve upon the seasonal climatological baseline at longer forecast horizons.

The maps in Figure~\ref{fig:maps_best_model_tw_pcrps_q099_vs_era5} display the best-performing model according to the $\overline{\operatorname{twPCRPS}}_t$ when forecasting exceedances of historical 99th percentile thresholds. As discussed in Section \ref{sec:results}, this comparison does not account for whether the differences in predictive performance between models are statistically significant. To account for this, we implement a block permutation test to assess whether the difference in $\overline{\operatorname{twPCRPS}}_t$ between all pairs of models is significant; this is the same test as that implemented by \cite{gneiting2026pcrps}. 

\begin{figure}[t!]
	\centering
	\includegraphics[width=\linewidth]{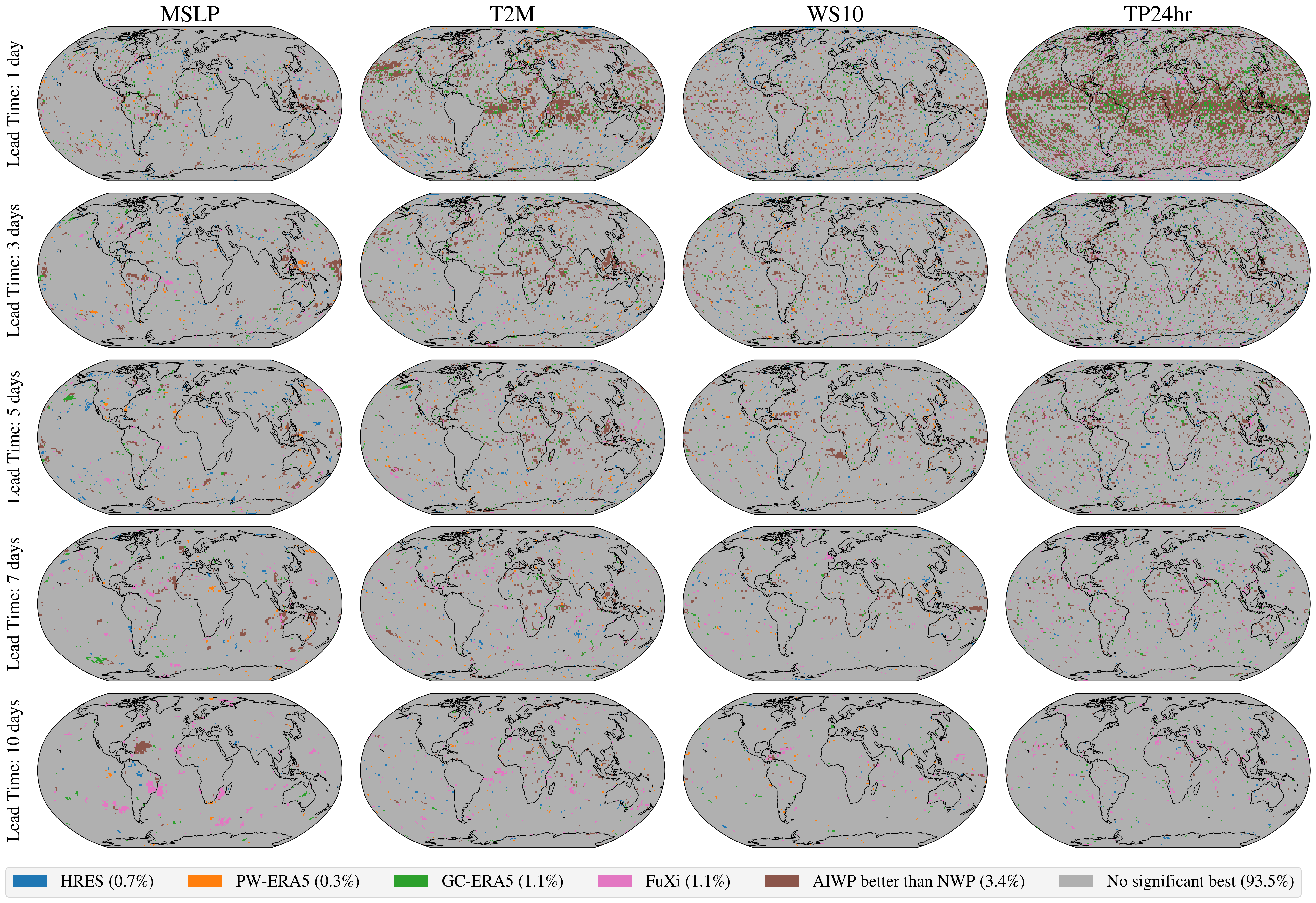}
	\caption{Significantly best-performing model at each grid point according to the $\overline{\operatorname{twPCRPS}}_t$ when predicting exceedances of the historical 99th percentile. At each grid point, the colour indicates whether a model receives a $\overline{\operatorname{twPCRPS}}_t$ that is significantly lower than that of all other available models at the 5\% significance level. Columns correspond to weather variables and rows correspond to lead times. The percentages in the legend correspond to the total proportion of cases where each model performs significantly better than all competitors, aggregated across all grid points, variables, and lead times. Grid points for which all available \gls{AIWP} models significantly outperform HRES, but no individual \gls{AIWP} model is significantly best, are also indicated. Grey grid points indicate locations where no significantly best model is identified. PW-ERA5 is not available for TP24hr and is therefore excluded from that column. ERA5 reanalyses are used as verification data.}
	\label{fig:maps_sig_best_model_ai_tw_pcrps_q099_vs_era5}
\end{figure}

Figure~\ref{fig:maps_sig_best_model_ai_tw_pcrps_q099_vs_era5} displays the significantly best-performing model at each grid point when forecasting exceedances of the historical 99th percentile. A model is ``significantly best-performing'' if it is found to significantly outperform all other models. We additionally include a category whereby none of the models significantly outperform all others but every available \gls{AIWP} model significantly outperforms the HRES model. Since we test for significant differences between all pairs of models, we are simultaneously performing multiple hypothesis tests, which raises the issue of multiple testing. This is not directly accounted for here, but since Figure~\ref{fig:maps_sig_best_model_ai_tw_pcrps_q099_vs_era5} displays cases where a model outperforms \emph{all} other models rather than \emph{any} of the other models, the results will be conservative; if that model does not truly outperform all competitors, then the probability of incorrectly declaring it significantly best-performing is no greater than the nominal significance level (in this case 0.05).

The results in Figure~\ref{fig:maps_sig_best_model_ai_tw_pcrps_q099_vs_era5} show that a significantly best-performing model is rarely identified, with 93.5\% of cases classified as ``no significant best''. In a further 3.4\% of cases, all available \gls{AIWP} models significantly outperform HRES without any individual model being significantly best. These cases are most widespread at a lead time of one day for precipitation and temperature, particularly in the tropics. For precipitation, GraphCast is also significantly best at many tropical grid points at this lead time. At longer lead times, FuXi is more often significantly best, but this advantage is confined to relatively few locations.

\begin{figure}
	\centering
	\includegraphics[width=\linewidth]{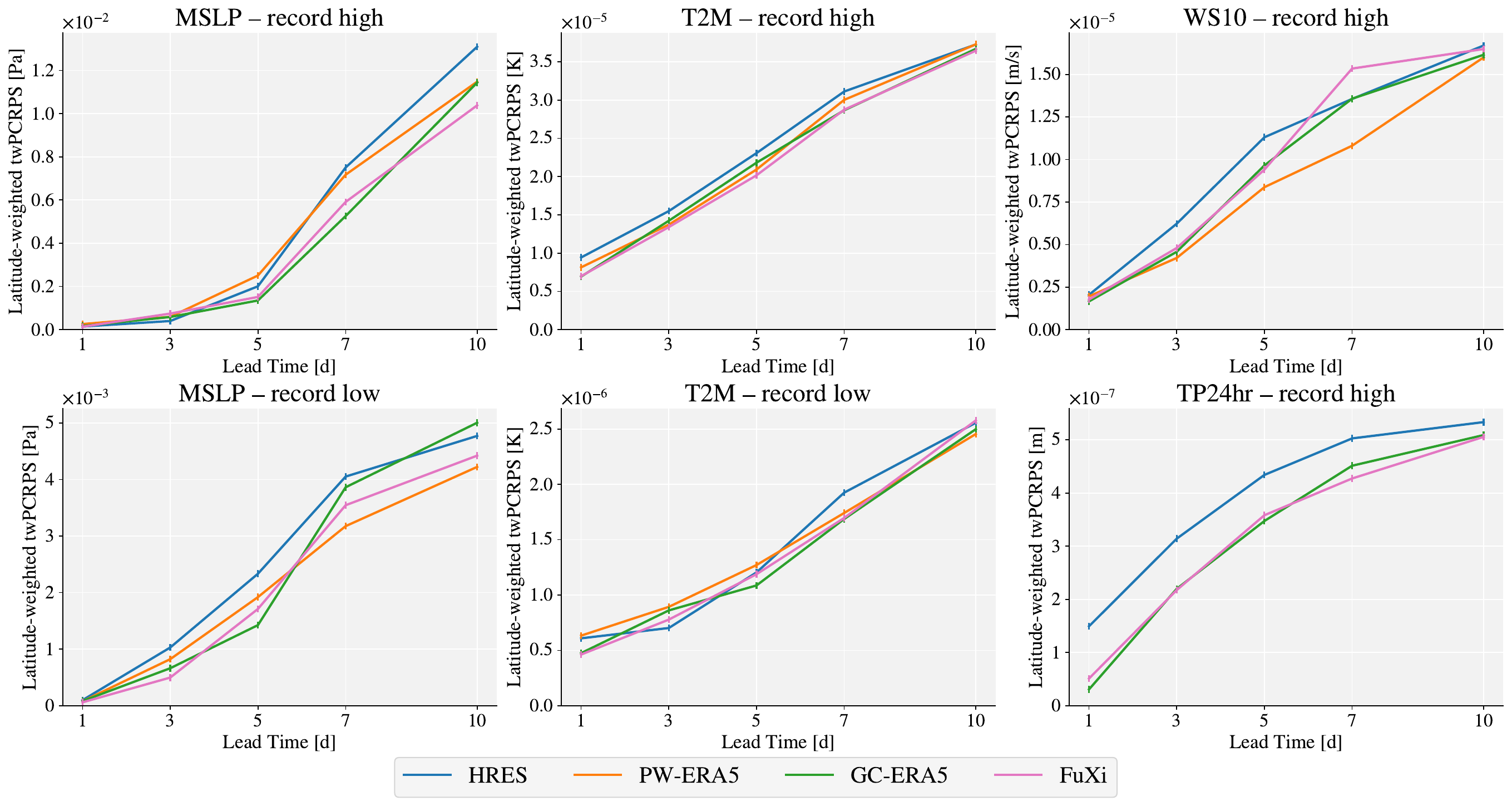}
	\caption{$\overline{\operatorname{twPCRPS}}_t$ for the four forecasting models when interest is on exceedances of overall record high values and non-exceedances of overall record low values. Results have been aggregated across all grid points and are displayed as a function of lead time. ERA5 reanalyses are used as verification data.}
	\label{fig:tw_pcrps_q100_vs_era5}
\end{figure}

\begin{figure}
	\centering
	\includegraphics[width=\linewidth]{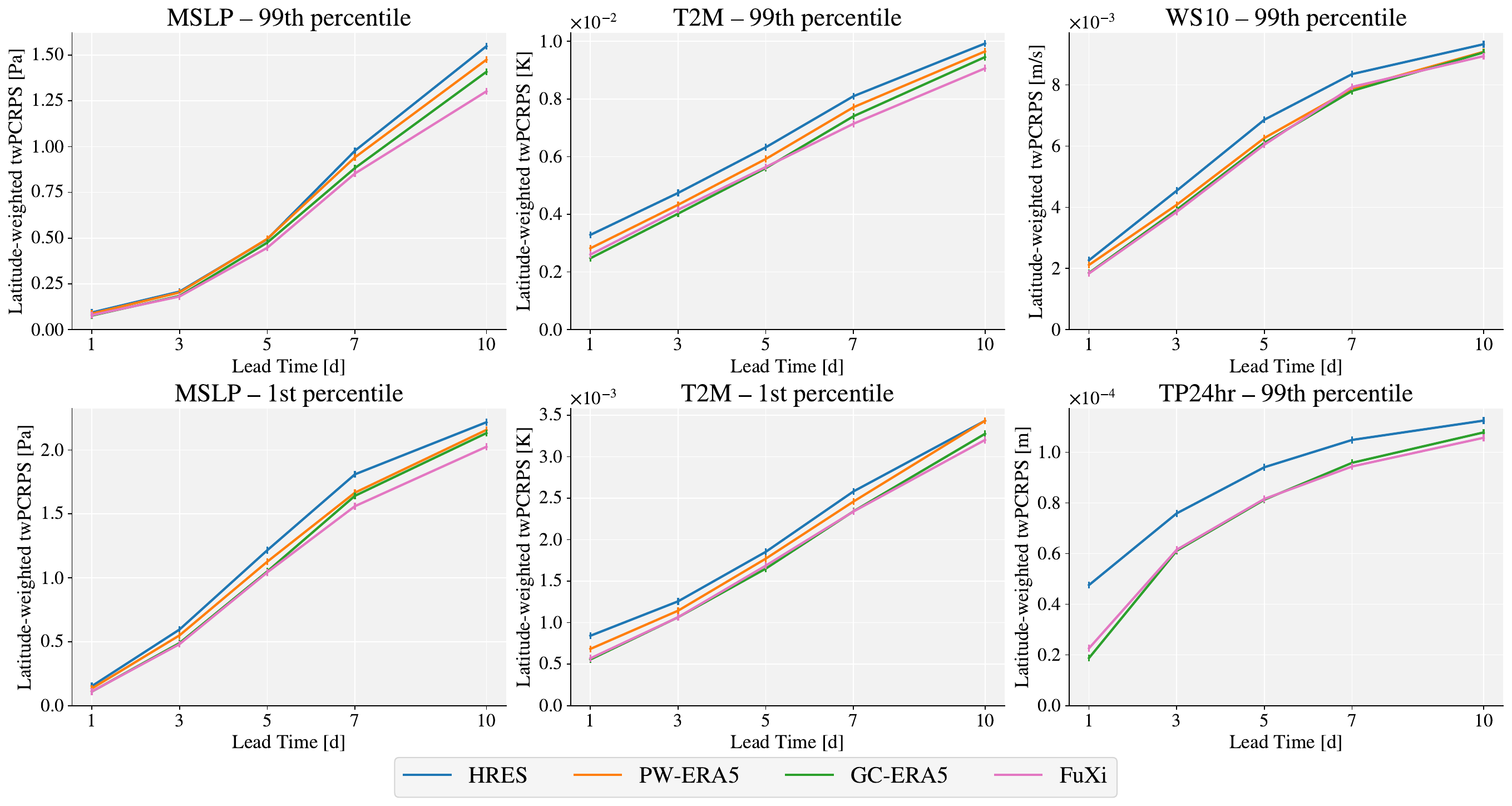}
	\caption{$\overline{\operatorname{twPCRPS}}_t$ for the four forecasting models when interest is on exceedances of the historical 99th percentile and non-exceedances of historical 1st percentile. Results have been aggregated across all grid points and are displayed as a function of lead time. ERA5 reanalyses are used as verification data.}
	\label{fig:tw_pcrps_q099_vs_era5}
\end{figure}

Figure~\ref{fig:tw_pcrps_q100_vs_era5} displays the $\overline{\operatorname{twPCRPS}}_t$ for the four models when forecasting overall record high and low events for each variable, with results aggregated over all grid points. Corresponding results for the historical 99th and 1st percentiles are shown in Figure~\ref{fig:tw_pcrps_q099_vs_era5}.
The results in both figures are broadly consistent with those for monthly records in Figure~\ref{fig:tw_pcrps_monthly_records_vs_era5}. 
In most cases, we observe that the \gls{AIWP} models have more potential than \gls{NWP} models to accurately predict extreme weather events. For overall records, Pangu-Weather is more competitive, achieving the lowest scores for wind speed from three days onwards and for record low MSLP at seven and ten days. For precipitation, GraphCast and FuXi generally achieve similar scores for overall records and the 99th percentile, whereas GraphCast consistently outperforms FuXi for monthly record thresholds.

Similar results are also obtained when the models are evaluated against IFS analysis fields instead of ERA5 reanalyses. In this operational comparison, we use HRES together with the IFS-initialised variants of GraphCast and Pangu-Weather, denoted by GC-IFS and PW-IFS, respectively. This avoids comparing HRES against \gls{AIWP} forecasts initialised from ERA5, which are not available in real time and may therefore give \gls{AIWP} models an advantage when ERA5 is also used as the verifying dataset. FuXi is not included because an IFS-initialised FuXi variant is not available in WeatherBench 2. Precipitation is also omitted since TP24hr is not available in the IFS analysis fields used as verification data. The evaluation year, 2020, falls within the 2016--2021 period used to fine-tune GC-IFS on HRES data.\footnote{See the documentation at \url{https://github.com/google-deepmind/weathernext/blob/main/docs/weathernext1_graph/README.md}.}

\begin{figure}
	\centering
	\includegraphics[width=\linewidth]{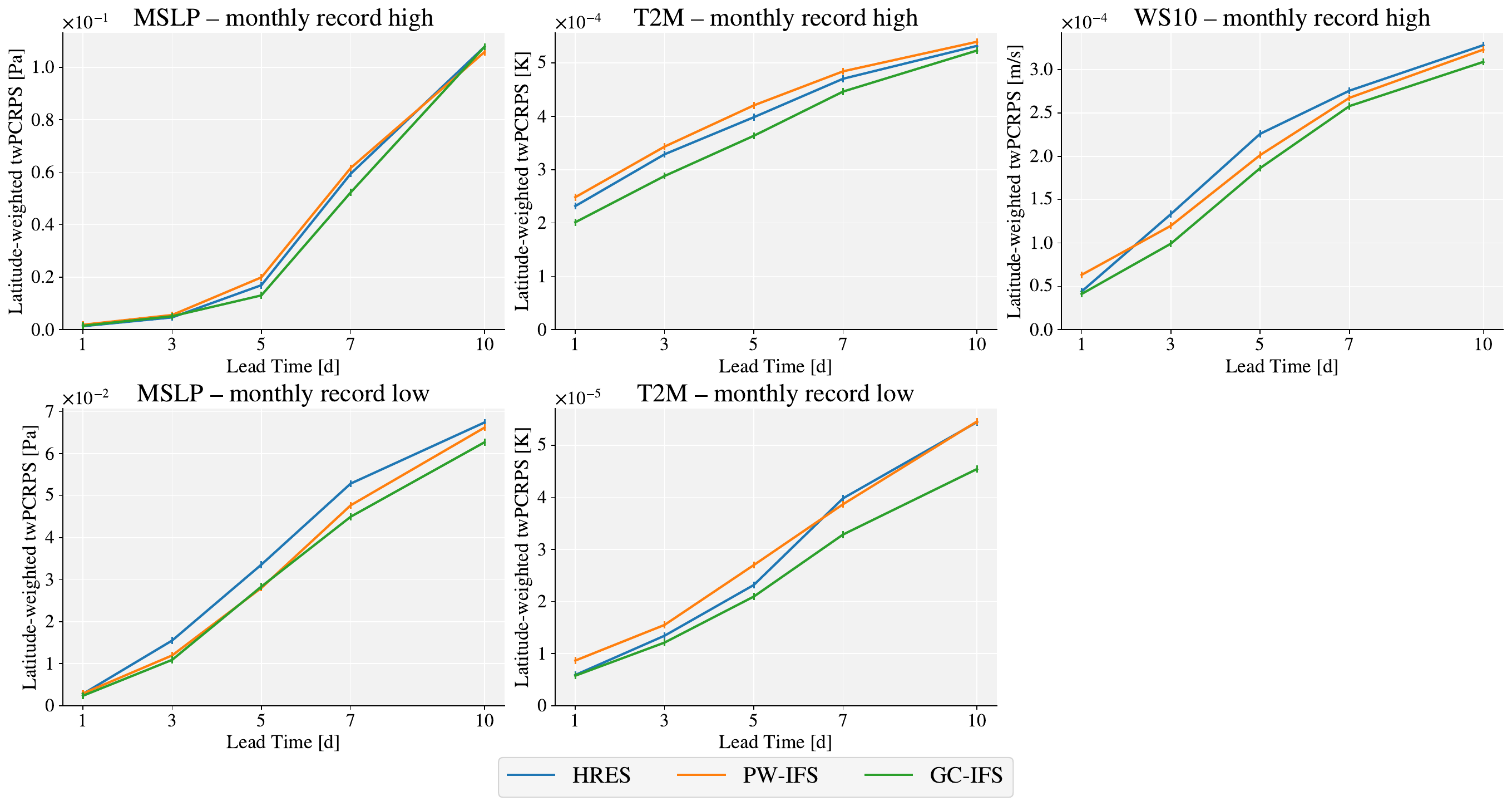}
	\caption{$\overline{\operatorname{twPCRPS}}_t$ for HRES and the IFS-initialised AIWP variants GC-IFS and PW-IFS, when interest is on exceedances of monthly record high values and non-exceedances of monthly record low values. The records are calculated using ERA5 reanalysis data, since this is available for a longer historical period than the IFS analyses. Scores are averaged across all grid points using a latitude weighting, and are shown as a function of lead time. IFS analyses are used as verification data.}
	\label{fig:tw_pcrps_monthly_records_operational}
\end{figure}

Figure~\ref{fig:tw_pcrps_monthly_records_operational} illustrates that GC-IFS generally outperforms HRES when forecasting monthly record-breaking events for the available variables, whereas PW-IFS achieves lower scores than HRES mainly for record lows of MSLP and record-breaking wind speeds. This is broadly consistent with the comparison in Figure~\ref{fig:tw_pcrps_monthly_records_vs_era5}, where GC-ERA5 also outperforms HRES and PW-ERA5 is particularly competitive for the same two record types.

\begin{figure}
	\centering
	\includegraphics[width=0.77\linewidth,keepaspectratio]{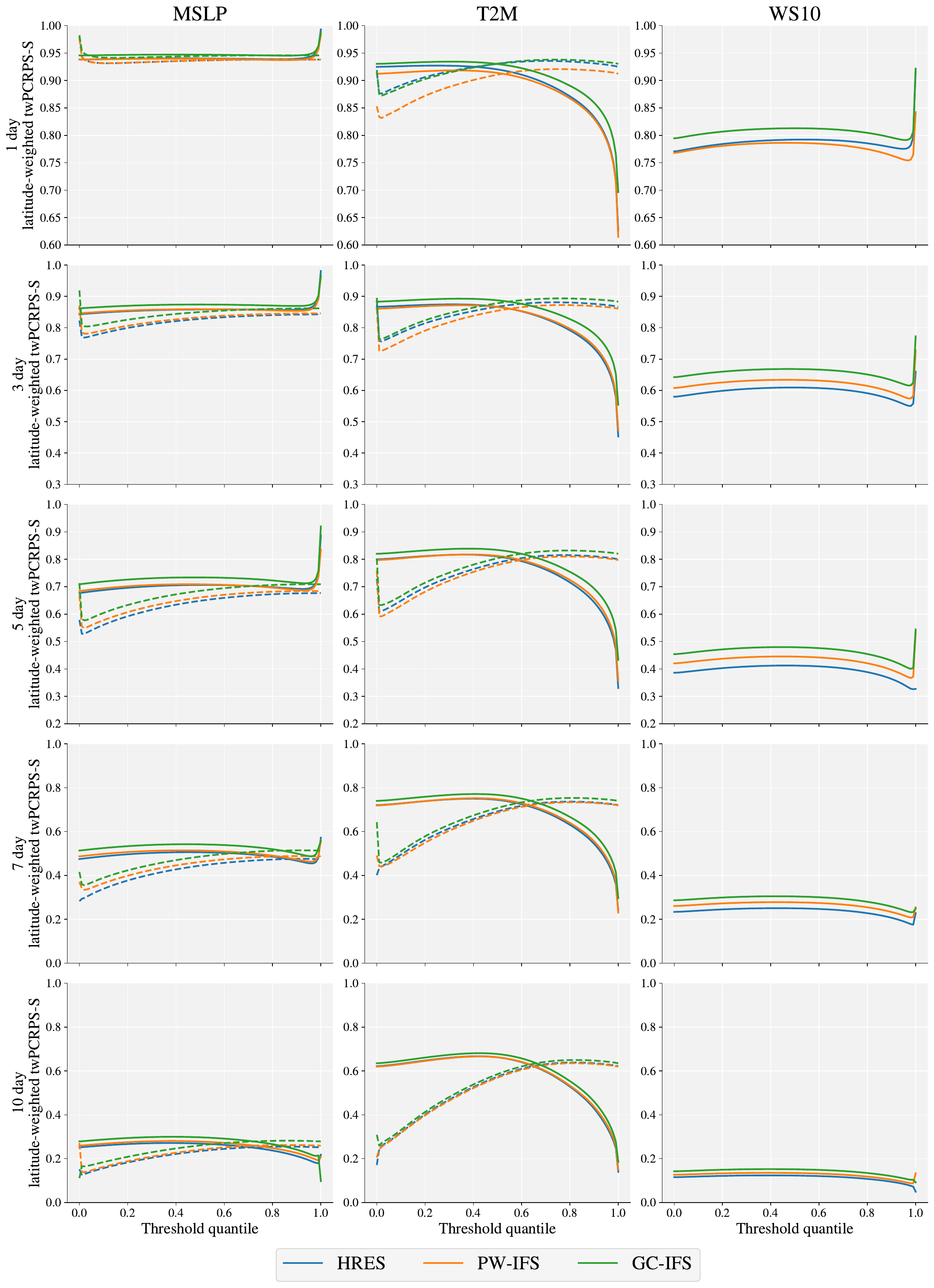}
	\caption{Latitude-weighted $\operatorname{twPCRPS-S}_t$ for the three operational forecasting models as a function of the quantile of the historical data used to define the threshold in the $\operatorname{twCRPS}_t$. Solid lines correspond to threshold exceedances. Dashed lines correspond to threshold non-exceedances and are shown for MSLP and T2M. Columns correspond to weather variables, and rows to lead times. Results are aggregated across all grid points, using IFS analyses as verification data. Note that the y-axis scale is the same for all variables but differs across lead times.}
	\label{fig:tw_pcrpss_operational}
\end{figure}

Figure~\ref{fig:tw_pcrpss_operational} extends the operational comparison to the historical quantile thresholds used in Figure~\ref{fig:tw_pcrpss_vs_era5}. GC-IFS generally achieves the highest skill across thresholds. PW-IFS and HRES have similar skill for MSLP and temperature at most thresholds. For temperature at shorter lead times, the skill gap between HRES and GraphCast is smaller in the operational comparison, particularly for threshold non-exceedances. In both comparisons, skill declines markedly for all models as the temperature exceedance threshold approaches the highest quantiles. For wind speed, PW-IFS is generally more skilful than HRES for lead times longer than 1~day.

\begin{figure}
	\centering
	\includegraphics[width=\linewidth]{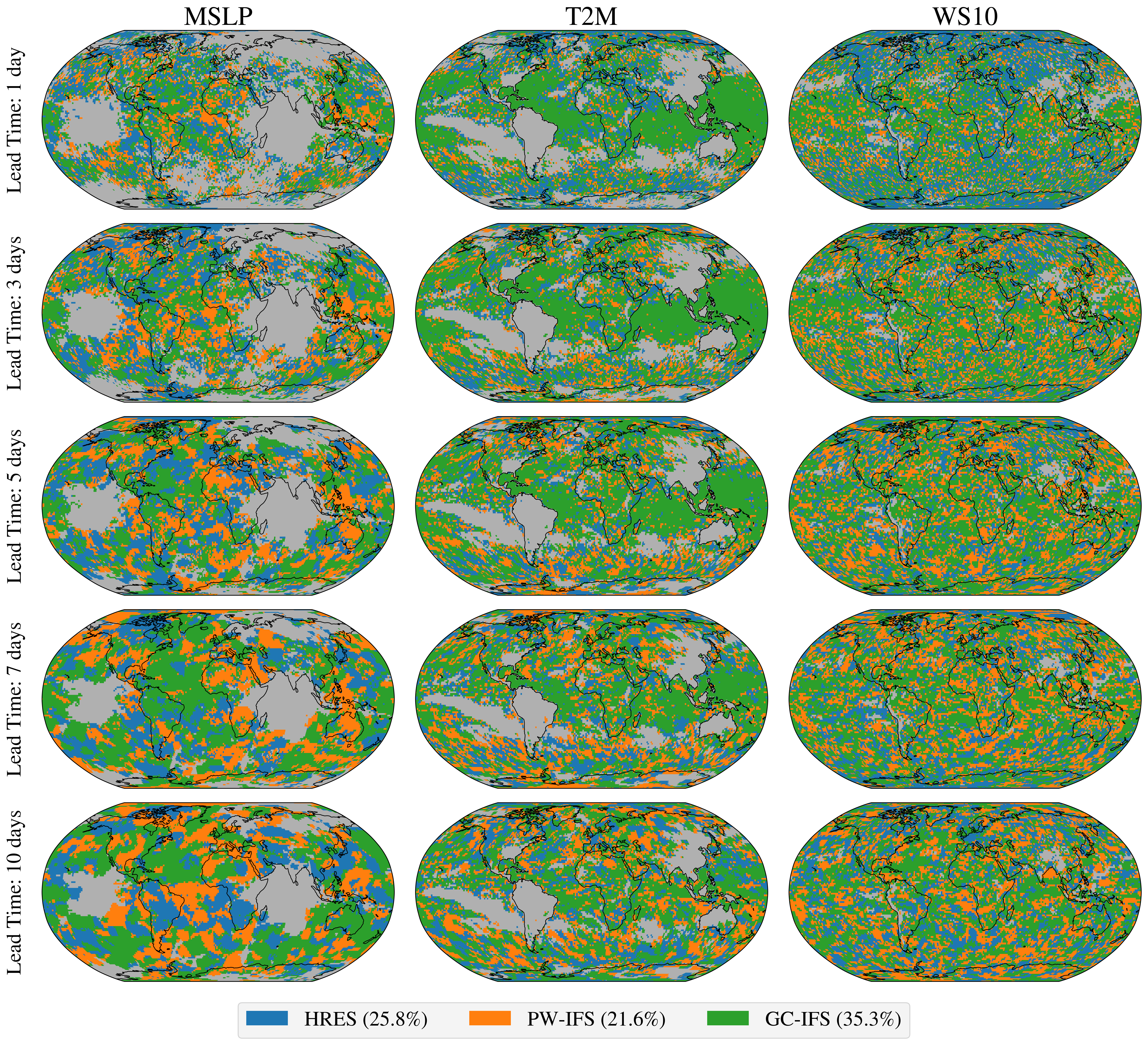}
	\caption{Best-performing model at each grid point according to $\overline{\operatorname{twPCRPS}}_t$ when predicting exceedances of the historical 99th percentile. At each grid point, the colour indicates the model with the lowest $\overline{\operatorname{twPCRPS}}_t$. Columns correspond to weather variables and rows to lead times. The percentages in the legend give the proportion of cases where each model performs best, aggregated across all grid points, variables, and lead times. Grey grid points indicate ties between two or more models, which primarily occur when multiple models attain a score of zero; these are not assigned to any individual model in the legend percentages. IFS analyses are used as verification data.}
	\label{fig:maps_best_operational_model_tw_pcrps_q099_vs_ifs_analysis}
\end{figure}

Figure~\ref{fig:maps_best_operational_model_tw_pcrps_q099_vs_ifs_analysis} shows the best-performing model at each grid point for exceedances of the historical 99th percentile, analogous to Figure~\ref{fig:maps_best_model_tw_pcrps_q099_vs_era5}.
GC-IFS has the unique lowest score most often (35.3\% of all grid point, variable, and lead time combinations) and does so over large areas for T2M.
The best-performing model varies more across locations for MSLP and WS10.

\begin{figure}
	\centering
	\includegraphics[width=\linewidth]{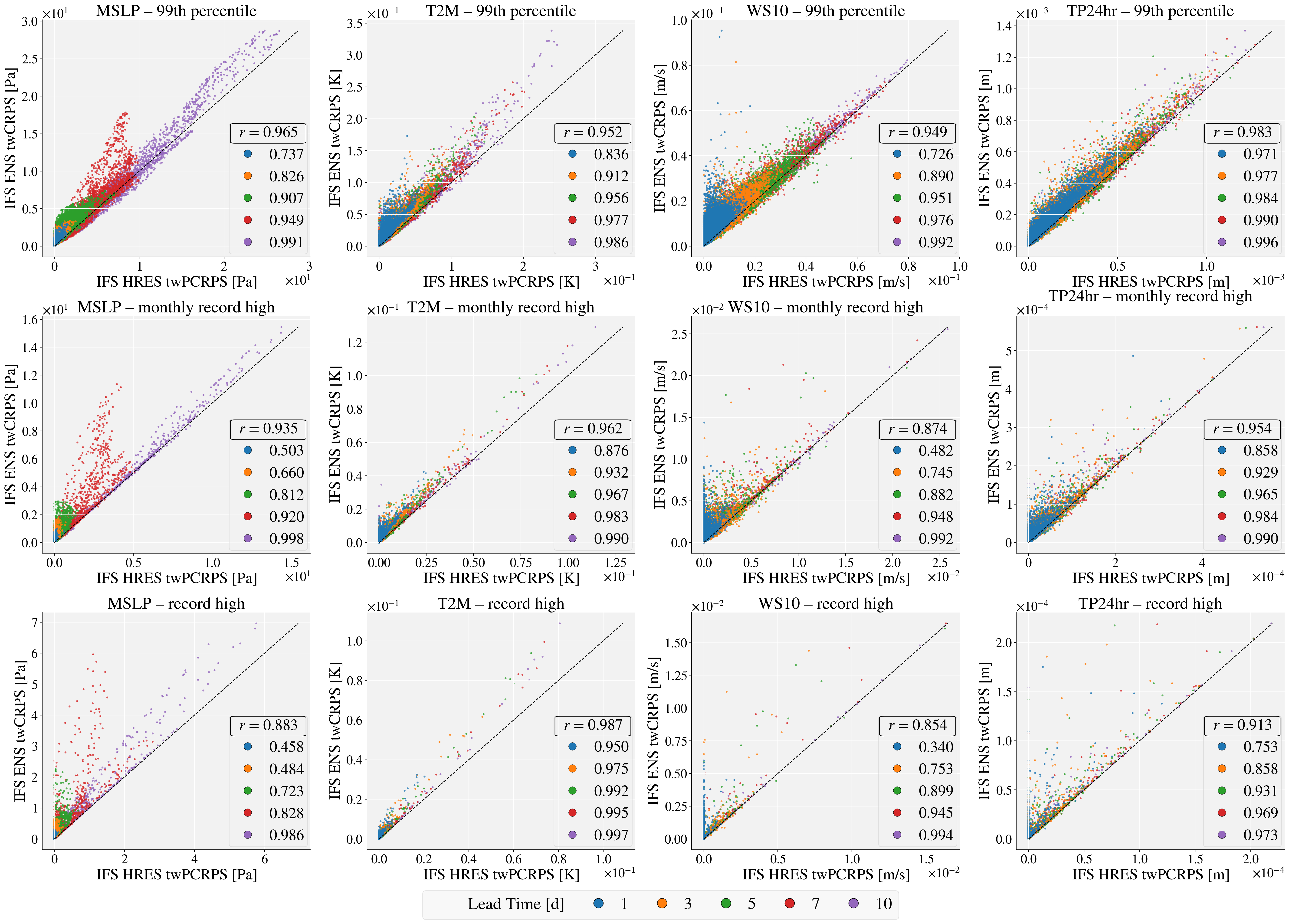}
	\caption{Scatterplots comparing the $\operatorname{twCRPS}_t$ of the IFS ensemble with the $\operatorname{twPCRPS}_t$ of the deterministic HRES forecast. Columns correspond to weather variables and rows to thresholds used in the evaluation: the historical 99th percentile (top), the monthly record threshold (middle), and the overall record threshold (bottom). Points correspond to grid point and lead time combinations, with colours indicating lead time. The dashed diagonal indicates equality between $\operatorname{twPCRPS}_t$ and $\operatorname{twCRPS}_t$, and the inset values report Pearson correlations overall and separately by lead time. ERA5 reanalyses are used as verification data.
	}
	\label{fig:scatterplot_nwp_tw_crps_vs_era5}
\end{figure}

Figure~\ref{fig:scatterplot_nwp_tw_crps_vs_era5} demonstrates how the \gls{twPCRPS} of the deterministic HRES forecast compares to the \gls{twCRPS} of the operational IFS ensemble forecasts. Results are shown separately for all lead times and weather variables, and for when interest is on exceedances of the 99th percentile, monthly records, and overall records. As for GraphCast and GenCast in Figure~\ref{fig:scatterplot_aiwp_tw_crps_vs_era5}, there is a strong positive correlation between the potential accuracy of the deterministic model and the actual accuracy of the probabilistic model when forecasting extreme events, for all variables. The correlation is again largest at longer lead times, with the correlation at 10 days exceeding 0.97 for all variables and considered thresholds.

To assess how much of the potential predictive performance can be realised by post-processing forecasts with EasyUQ fitted to separate training data, we compare in-sample twPCRPS and out-of-sample twCRPS on the same 2020 evaluation cases. We use forecasts initialised in the two years preceding the evaluation, 2018 and 2019, and their corresponding ERA5 observations to train EasyUQ. We restrict the comparison to PW-ERA5 and HRES because they are the only models in the main comparison for which WeatherBench~2 provides forecasts throughout both years. We consider MSLP, T2M, and WS10, the three study variables available for both models, at the same five lead times as in the main comparison. Fits are obtained separately for each model, variable, grid point, and lead time, using only training forecasts with valid times before 1~January 2020. We apply the fitted mappings to the 702 evaluation cases per lead time and evaluate the resulting predictive distributions against ERA5.

\begin{figure}
	\centering
	\includegraphics[width=\linewidth]{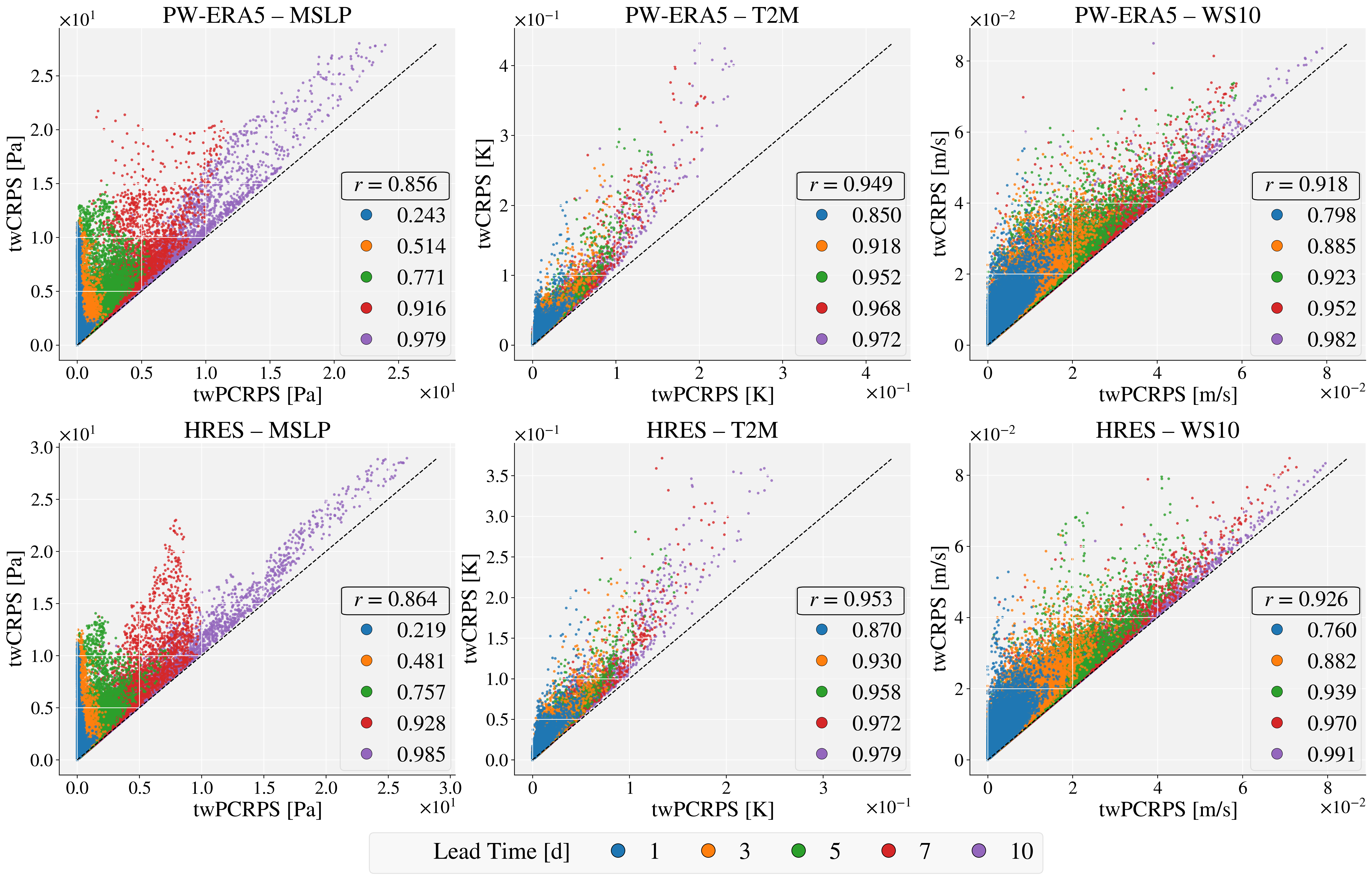}
	\caption{Scatterplots comparing the in-sample $\overline{\operatorname{twPCRPS}}_t$ and out-of-sample $\overline{\operatorname{twCRPS}}_t$ for PW-ERA5 and HRES when forecasting exceedances of the historical 99th percentile. Rows correspond to models and columns to weather variables. Points correspond to grid point and lead time combinations, with colours indicating lead time. The dashed diagonal indicates equality, and the inset values report Pearson correlations overall and separately by lead time. ERA5 reanalyses are used as verification data.}
	\label{fig:scatterplot_oos_tw_crps_q099_vs_era5}
\end{figure}

Figure~\ref{fig:scatterplot_oos_tw_crps_q099_vs_era5} compares the in-sample and out-of-sample scores for exceedances of the historical 99th percentile. The points lie on or above the diagonal, as implied by the lower bound in Appendix~A. The Pearson correlations between the two scores increase with lead time for both models and all three variables. This increase is particularly pronounced for MSLP, for which the Pearson correlations rise from 0.24 and 0.22 at 1~day to 0.98 and 0.99 at 10~days for PW-ERA5 and HRES, respectively. For T2M and WS10, the two scores are already strongly correlated at 1~day.

\begin{figure}
	\centering
	\includegraphics[width=\linewidth]{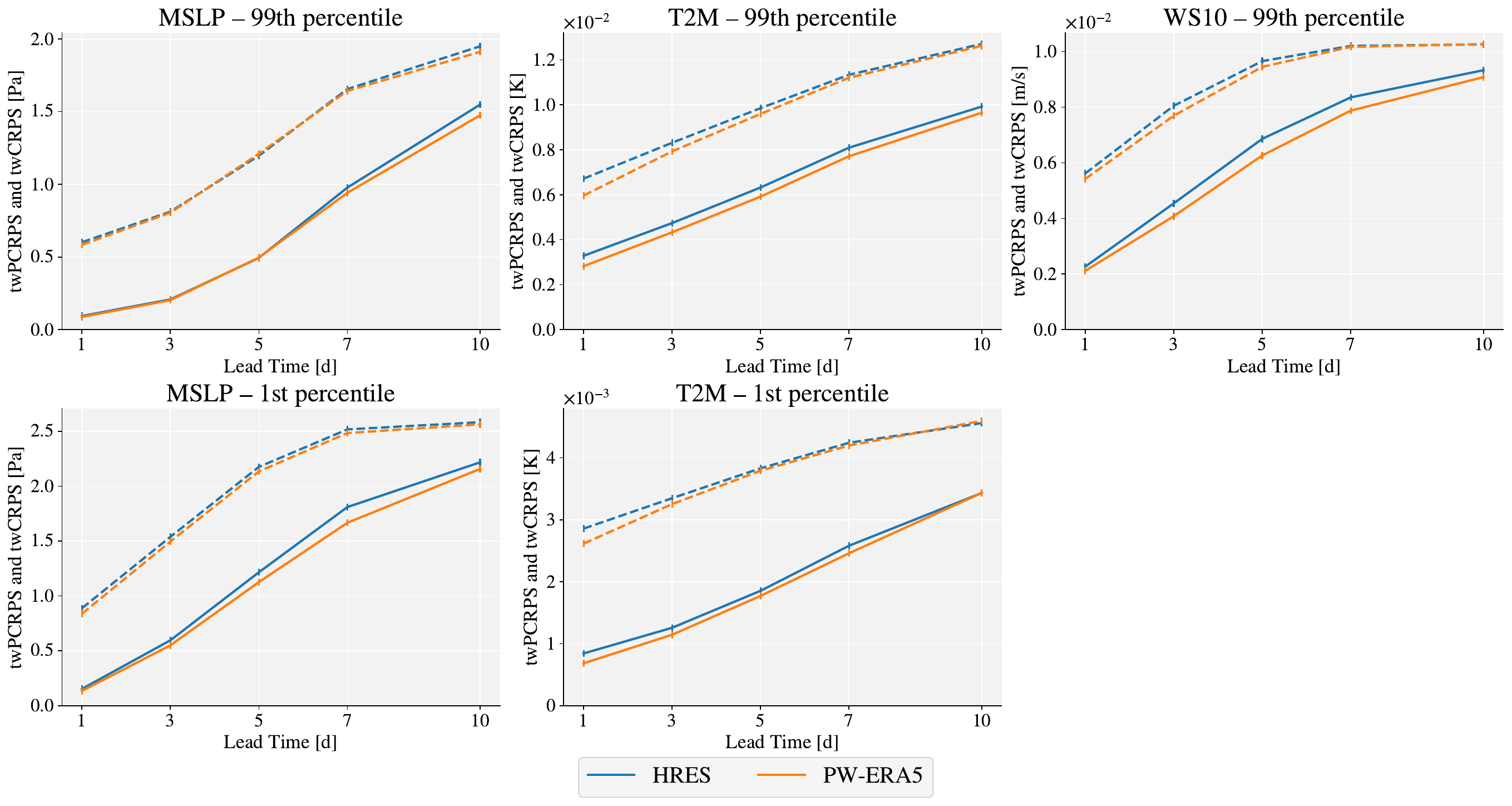}
	\caption{Scores for PW-ERA5 and HRES when interest is on exceedances of the historical 99th percentile and non-exceedances of the historical 1st percentile. Solid lines show the in-sample $\overline{\operatorname{twPCRPS}}_t$ and dashed lines show the out-of-sample $\overline{\operatorname{twCRPS}}_t$. Results have been aggregated over the same forecast cases and across all grid points using a latitude weighting. ERA5 reanalyses are used as verification data.}
	\label{fig:tw_pcrps_q099_vs_era5_oos}
\end{figure}

Figure~\ref{fig:tw_pcrps_q099_vs_era5_oos} shows the latitude-weighted scores as a function of lead time for exceedances of the historical 99th percentile and non-exceedances of the historical 1st percentile. For both tails, the absolute increases in score out of sample are similar for PW-ERA5 and HRES and exceed the differences between the two models. Relative to the in-sample scores, these increases are largest at 1~day, particularly for MSLP, for which the out-of-sample scores at this lead time are roughly six times the in-sample scores for both tails. The model ranking is largely preserved out of sample, with PW-ERA5 generally achieving the lower score.

\begin{figure}
	\centering
	\includegraphics[width=0.77\linewidth,keepaspectratio]{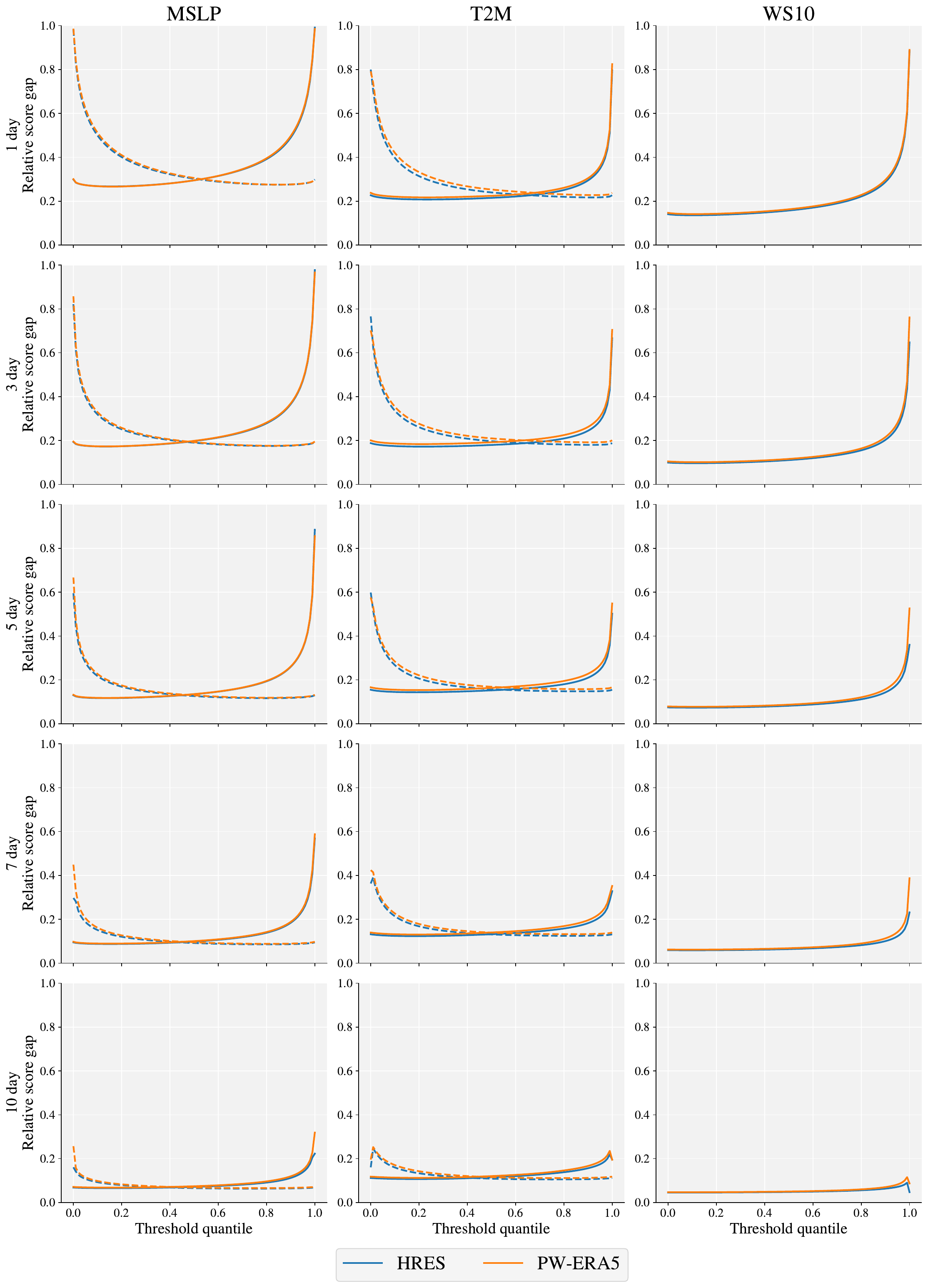}
	\caption{Relative gap between the latitude-weighted in-sample $\overline{\operatorname{twPCRPS}}_t$ and out-of-sample $\overline{\operatorname{twCRPS}}_t$ for PW-ERA5 and HRES as a function of the historical threshold quantile. A value of zero indicates equal scores, and larger values indicate a wider gap relative to the out-of-sample score. Solid lines correspond to threshold exceedances. Dashed lines correspond to threshold non-exceedances and are shown for MSLP and T2M. Columns correspond to weather variables and rows to lead times. ERA5 reanalyses are used as verification data.}
	\label{fig:tw_pcrpss_vs_era5_oos}
\end{figure}

Figure~\ref{fig:tw_pcrpss_vs_era5_oos} shows how the relative score gap varies with the threshold quantile. For each model, we calculate this gap as one minus the ratio of the in-sample twPCRPS to the out-of-sample twCRPS, using the latitude-weighted mean of each score. The gap generally widens towards more extreme thresholds in each tail, especially at short lead times. At 1~day, it approaches one at both overall MSLP record thresholds. PW-ERA5 and HRES show similar relative gaps across most thresholds, with the clearest differences near the overall records.

For monthly and overall records, the training observations are part of the 1979--2019 ERA5 archive used to define the records. The out-of-sample scores therefore reduce to the observed record departures in Equation~\eqref{eq:twcrps_oos_records}, averaged over all verification cases. This yields identical scores for PW-ERA5 and HRES at each grid point and lead time.

\section{Quantile-weighted PCRPS}\label{app:qwpcrps}

\subsection{Definition}

The quantile-weighted $\operatorname{CRPS}$ \citep{gneiting2011wcrps} is defined as
\[
\operatorname{qwCRPS}_w(F,y) = 2 \int_0^1 \left(\mathbbm{1}_{\{y\leq F^{-1}(\alpha)\}}-\alpha\right)\left(F^{-1}(\alpha)-y\right) w(\alpha) d\alpha,
\]
where $F^{-1}$ is the quantile function associated with $F$, and $w$ is a non-negative weight function on the unit interval that can be chosen to emphasise different quantile levels of the forecast distribution. The CRPS is recovered when $w(\alpha) = 1$ for all $\alpha \in (0, 1)$ \citep{laio2007verification}.

To focus on the tails of the predictive distribution, common weight functions include $w(\alpha) = \mathbbm{1}_{\{\alpha > \tau\}}$ or $w(\alpha) = \mathbbm{1}_{\{\alpha < \tau\}}$, depending on whether interest is on the upper or lower tail of the predictive distribution, for some threshold $\tau \in (0, 1)$. If interest is on the upper tail of the forecast distribution, this yields
\[
\operatorname{qwCRPS}_\tau(F,y) = 2 \int_\tau^1 \left(\mathbbm{1}_{\{y\leq F^{-1}(\alpha)\}}-\alpha\right)\left(F^{-1}(\alpha)-y\right) d\alpha,
\]
which demonstrates that the evaluation only concerns the forecast distribution at quantile levels greater than $\tau$. An analogous expression exists for the lower tail, with the integration restricted to $(0, \tau)$. 

In contrast to the twCRPS, the qwCRPS emphasises regions of the forecast distribution, rather than regions of the outcome space. Hence, even if the threshold $\tau$ is close to one, the qwCRPS may not directly target extreme events; a high threshold of a non-extreme forecast distribution does not correspond to an extreme outcome. For this reason, the twCRPS is generally preferred to the qwCRPS when evaluating forecasts for extreme events in practical applications.

Nonetheless, we can similarly construct a potential qwCRPS. Given a deterministic forecast $x$ with corresponding EasyUQ predictive distribution $\hat{F}$, the quantile-weighted PCRPS is defined as
\[
\operatorname{qwPCRPS}_\tau(x, y) = \operatorname{qwCRPS}_\tau(\hat{F}, y).
\]
A subscript $\tau$ is similarly used to clarify that this is a function of the threshold $\tau$. The forecasts $x_1, \dots, x_n$ can again be evaluated using the average $\operatorname{qwPCRPS}_\tau$ over all $n$ forecast cases, 
\[
\overline{\operatorname{qwPCRPS}}_\tau = \frac{1}{n}\sum_{i=1}^{n} \operatorname{qwPCRPS}_\tau(x_i,y_i) = \frac{1}{n}\sum_{i=1}^{n} \operatorname{qwCRPS}_\tau(\hat F_i,y_i).
\]

As with the CRPS and twCRPS, \citet[Theorem 2]{henzi2021isotonic} similarly show that the EasyUQ predictive distributions result in the optimal quantile-weighted CRPS over the test data. We therefore again argue that evaluating the resulting forecast distributions using the quantile-weighted CRPS does not favour any particular forecast model, providing an objective and fair means to compare \gls{AIWP} and \gls{NWP} models when focus is on the tails of the forecast distribution.

A skill score based on the qwPCRPS can be defined as before. The reference quantile-weighted CRPS is
%The same procedure applies to the baseline $\operatorname{twPCRPS^{(0)}}$, yielding
\[
\overline{\operatorname{qwPCRPS}}_{0,\tau} =\frac{1}{n} \sum_{i=1}^n \operatorname{qwCRPS}_\tau(F^0, y_i),
\]
where $F^0$ is again the empirical distribution of $y_1, \dots, y_n$. We discuss how to calculate the quantile-weighted CRPS for (weighted) empirical distributions in the following subsection. The corresponding skill score is defined as
\[
\operatorname{qwPCRPS-S}_\tau = 1 - \frac{\overline{\operatorname{qwPCRPS}}_\tau}{\overline{\operatorname{qwPCRPS}}_{0,\tau}}.
\]
The $\operatorname{qwPCRPS-S}_\tau$ is bounded between zero and one, and can be interpreted analogously to the $\operatorname{PCRPS-S}$ and $\operatorname{twPCRPS-S}_t$.

\begin{figure}
	\centering
	\includegraphics[width=\linewidth]{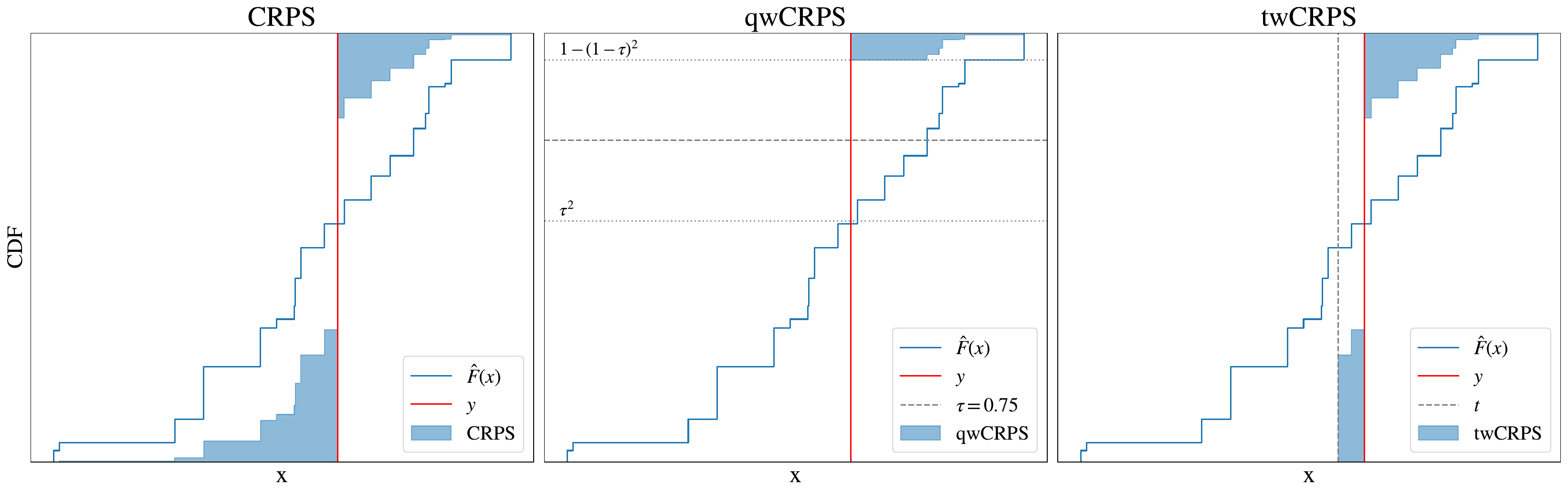}
	\caption{Illustration of the $\operatorname{CRPS}$ (left), $\operatorname{qwCRPS}$ (centre), and $\operatorname{twCRPS}$ (right) of an EasyUQ predictive distribution $\hat F$ and observation $y$. The qwCRPS uses the weight function $w(\alpha) = \one_{\{\alpha > 0.75\}}$, with the dotted horizontal lines indicating the corresponding transformed probability levels. The twCRPS uses the weight function $w(z) = \one_{\{z > t\}}$.}
	\label{fig:weighted_crps}
\end{figure}

The difference between the CRPS, twCRPS, and qwCRPS is displayed graphically in Figure \ref{fig:weighted_crps} for an EasyUQ predictive distribution when predicting T2M. The CRPS integrates the squared difference between the forecast distribution $\hat{F}$ and the step function defined by the observation $y$ across the $x$-axis. The squared difference at each temperature is shown by the shaded region in the first panel of Figure \ref{fig:weighted_crps}. The twCRPS restricts the score to the squared distance integrated across all temperatures greater than the chosen threshold, resulting in the smaller shaded region in the third panel of Figure \ref{fig:weighted_crps}. The qwCRPS instead restricts the score to the distance between $\hat{F}$ and $y$ above the chosen quantile threshold, as seen in the second panel of Figure \ref{fig:weighted_crps}.

\subsection{Computation}

Suppose we wish to calculate the qwCRPS for an EasyUQ predictive CDF $\hat F$, defined by support points $\hat{x}_1, \dots, \hat{x}_M$ with corresponding jumps in the distribution function of size $\omega_1, \dots, \omega_M$, with $\sum_{m=1}^M \omega_m = 1$. This can be interpreted as a weighted ensemble forecast, with ensemble members $\hat{x}_1, \dots, \hat{x}_M$ and weights $\omega_1, \dots, \omega_M$. We assume without loss of generality that the support points are ordered, $\hat{x}_1 \le \dots \le \hat{x}_M$, and we define $\tilde{\omega}_i = \sum_{m=1}^i \omega_m$ as the cumulative weight assigned to the first $i$ support points, for $i = 1, \dots, M$, with $\tilde{\omega}_0 = 0$. In this case, the quantile function of $\hat{F}$ is given by 
\[
\hat{F}^{-1}(\alpha) = \hat{x}_i \quad \text{for $\alpha \in (\tilde{\omega}_{i-1}, \tilde{\omega}_i]$}.
\]

Substituting this into the definition of the qwCRPS yields
\[
\operatorname{qwCRPS}_w(\hat{F},y) = 2 \sum_{m=1}^M \int_{\tilde{\omega}_{m-1}}^{\tilde{\omega}_m} (\one_{\{\hat{x}_m \ge y\}} - \alpha)(\hat{x}_m - y) w(\alpha) ~d\alpha.
\]
For an indicator weight function of the form $w(\alpha) = \one_{\{\alpha > \tau\}}$ for upper-tail emphasis, or $w(\alpha) = \one_{\{\alpha < \tau\}}$ for lower-tail emphasis, we obtain
\[
\operatorname{qwCRPS}_\tau(\hat{F},y) = 2 \sum_{m=1}^M \int_{\tilde{\omega}_{m-1}^*}^{\tilde{\omega}_m^*} (\one_{\{\hat{x}_m \ge y\}} - \alpha)(\hat{x}_m - y) ~d\alpha,
\]
where $\tilde \omega_m^\ast =\max\{\tilde \omega_m, \tau\}$ for upper-tail emphasis, and $\tilde \omega_m^\ast =\min\{\tilde \omega_m, \tau\}$ for lower-tail emphasis. This simplifies to
\[
\operatorname{qwCRPS}_\tau(\hat{F},y) = 2 \sum_{m=1}^M \left( \mathbbm{1}_{\{\hat{x}_m \ge y\}} \left(\tilde \omega_m^\ast - \tilde \omega_{m-1}^\ast \right) - \frac{1}{2} \left( \left(\tilde \omega_m^\ast \right)^2 - \left(\tilde \omega_{m-1}^\ast \right)^2 \right) \right) (\hat{x}_m-y).
\]

While we derive this expression to evaluate EasyUQ predictive distributions, it can more generally be used to calculate the quantile-weighted CRPS for any weighted ensemble forecast. When the weight function is equal to one (corresponding to $\tau = 0$ for upper-tail emphasis and $\tau = 1$ for lower-tail emphasis), we have $\tilde{\omega}_m^* = \tilde{\omega}_m$ for all $m$, and hence $\tilde{\omega}_m^* - \tilde{\omega}_{m-1}^* = \omega_m$ and $(\tilde{\omega}_m^*)^2 - (\tilde{\omega}_{m-1}^*)^2 = \tilde{\omega}_m^2 - \tilde{\omega}_{m-1}^2 = 2\omega_m \tilde{\omega}_m - \omega_m^2$, in which case the expression of the qwCRPS simplifies to
\[
\operatorname{qwCRPS}_\tau(\hat{F},y) = 2 \sum_{m=1}^M \omega_m \left( \mathbbm{1}_{\{\hat{x}_m \ge y\}} - \tilde{\omega}_m + \frac{\omega_m}{2} \right) (\hat{x}_m-y),
\]
which is the standard expression for the standard (unweighted) CRPS when the forecast is a weighted empirical distribution. If the weights of the ensemble members are all the same, $\omega_m = 1/M$, then $\tilde{\omega}_m = m/M$, and the expression simplifies further to
\[
\operatorname{qwCRPS}_\tau(\hat{F},y) = \frac{2}{M} \sum_{m=1}^M \left( \mathbbm{1}_{\{\hat{x}_m \ge y\}} - \frac{2m - 1}{2M} \right) (\hat{x}_m-y),
\]
which is the standard expression for the standard CRPS when the forecast is an unweighted empirical distribution \citep[see e.g.][Chapter 6]{jordan2016facets}.

This expression can also be used to calculate the qwCRPS for the unconditional climatological baseline, yielding
\begin{align*}
	\overline{\operatorname{qwPCRPS}}_{0,\tau} = \frac{2}{n} \sum_{i=1}^n \sum_{j=1}^n \left( \mathbbm{1}_{\{y_{(j)}\geq y_i\}} \left(\tilde \omega_j^\ast - \tilde \omega_{j-1}^\ast \right) - \frac{1}{2} \left( \left(\tilde \omega_j^\ast \right)^2 - \left(\tilde \omega_{j-1}^\ast \right)^2 \right) \right) (y_{(j)}-y_i),
\end{align*}
where $y_{(1)} \le y_{(2)} \le \dots \le y_{(n)}$ are the order statistics of the past observations, and $\tilde \omega_j^* = \max\{j/n, \tau\}$ or $\tilde \omega_j^* = \min\{j/n, \tau\}$, depending on whether interest is on the upper or lower tail, respectively.

\subsection{Results}

\begin{figure}
	\centering
	\includegraphics[width=\linewidth]{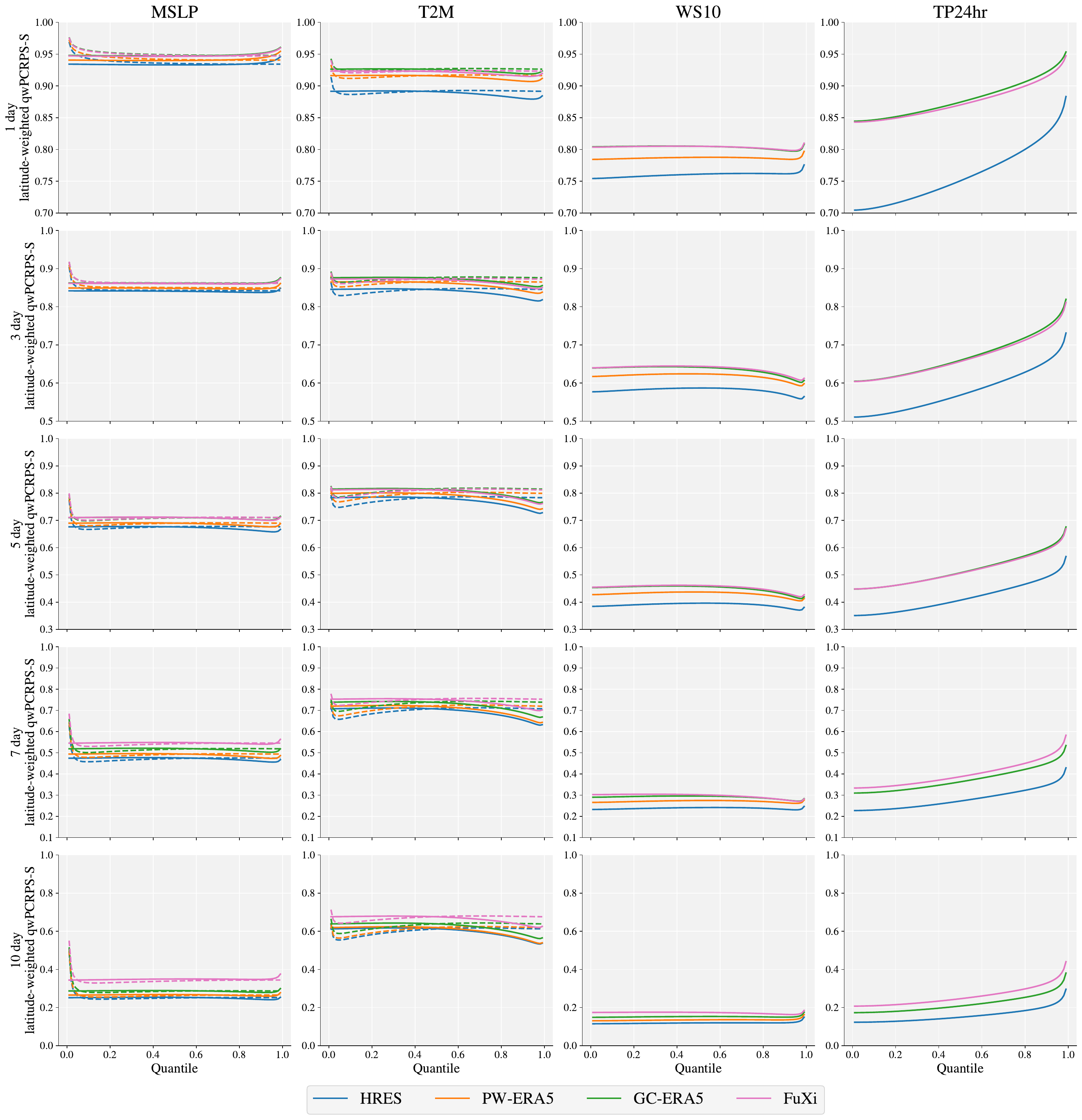}
	\caption{$\operatorname{qwPCRPS-S}_\tau$  for the four forecasting models as a function of the quantile level. Solid lines show upper-tail scores, dashed lines show lower-tail scores. Lower-tail scores are shown only for mean sea level pressure and 2~m temperature. The columns correspond to different weather variables, and the rows correspond to different lead times. Results are aggregated across all grid points, using ERA5 reanalyses as verification data. Note that the y-axis scale is the same for all variables but differs across lead times.}
	\label{fig:qw_pcrpss_vs_era5}
\end{figure}

Figure \ref{fig:qw_pcrpss_vs_era5} displays $\operatorname{qwPCRPS-S}_\tau$ for the four forecasting models as a function of the threshold $\tau$. Results are shown for each weather variable and lead time, for the 99 quantiles $\tau \in \{0.01, 0.02, \dots, 0.98, 0.99\}$. The results are qualitatively similar to those for the threshold-weighted PCRPS: while there are fluctuations in the skill scores themselves, the ordering of the competing models is generally insensitive to the threshold used within the quantile-weighted PCRPS. FuXi again tends to achieve the highest potential skill, particularly at longer lead times, while GraphCast is often competitive and occasionally performs best at shorter lead times; Pangu-Weather generally performs below GraphCast and FuXi, but above HRES.

\begin{figure}[t!]
	\centering
	\includegraphics[width=\linewidth]{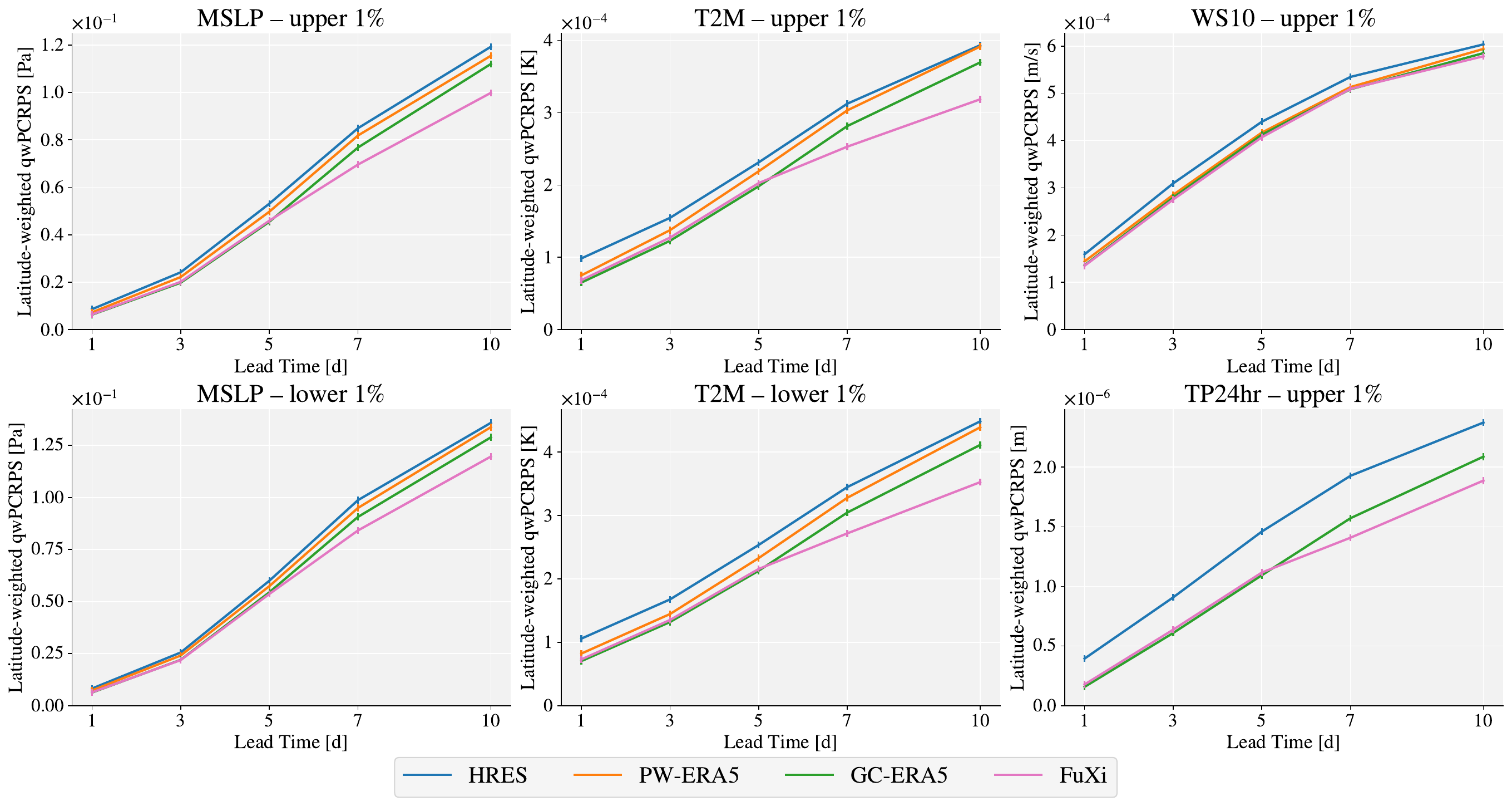}
	\caption{$\overline{\operatorname{qwPCRPS}}_\tau$ with weight function $w(\alpha) = \one_{\{\alpha > 0.99\}}$ (upper 1\%) and $w(\alpha) = \one_{\{\alpha < 0.01\}}$ (lower 1\%). Results have been aggregated across all grid points and are displayed as a function of lead time. ERA5 reanalyses are used as verification data.}
	\label{fig:qw_pcrps_99_vs_era5}
\end{figure}

Figure \ref{fig:qw_pcrps_99_vs_era5} shows the quantile-weighted PCRPS as a function of lead time, when evaluating the lower 1\% and the upper 1\% of the EasyUQ predictive distributions. We see an improvement in the relative performance of FuXi beyond lead times of five days, particularly for temperature, with the ordering of the forecasting methods the same regardless of whether focus is placed on the upper or lower tail.

\begin{figure}[t!]
	\centering
	\includegraphics[width=\linewidth]{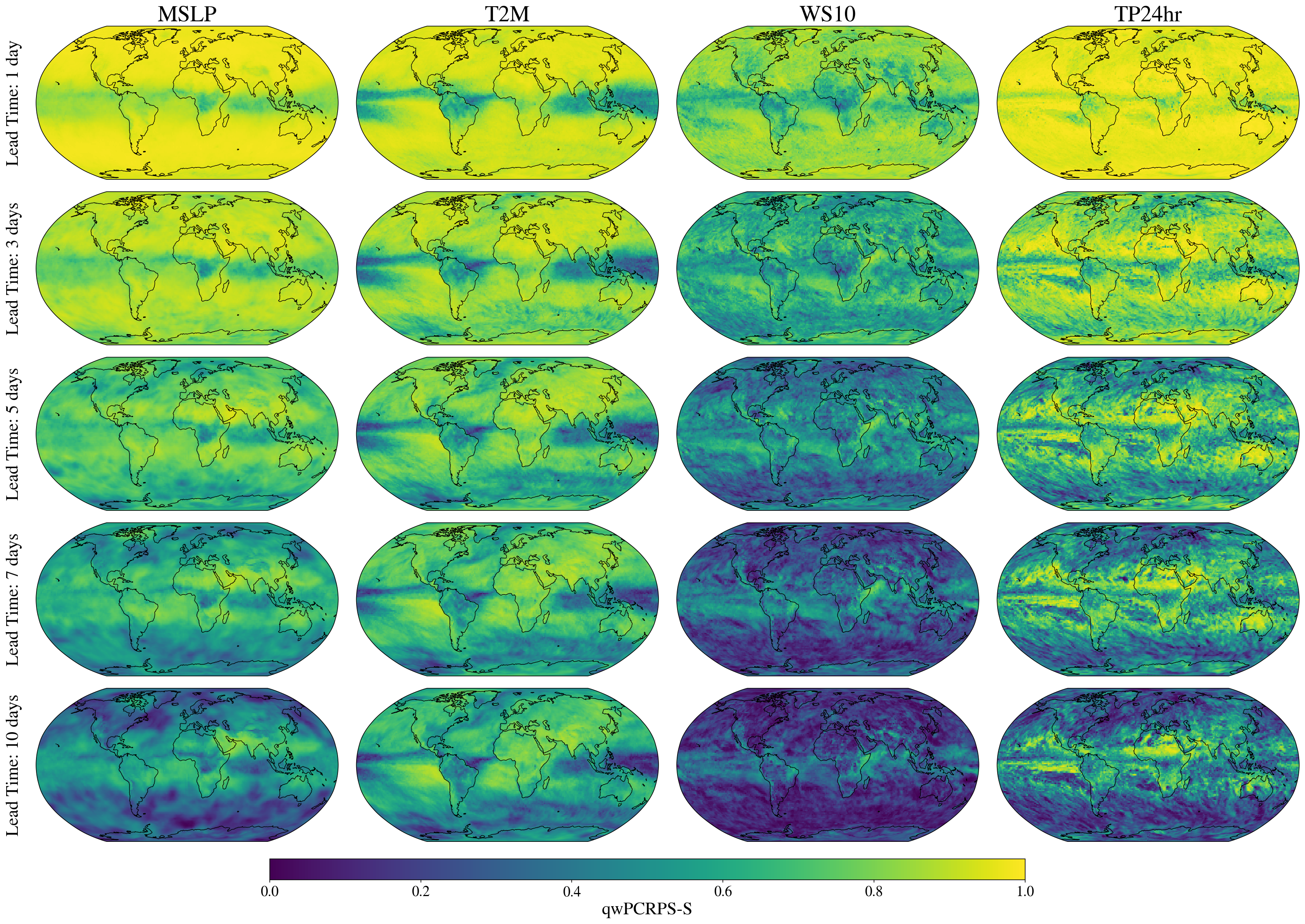}
	\caption{$\operatorname{qwPCRPS-S}_\tau$ of FuXi. Results are shown at each grid point using ERA5 reanalyses as verification data. The qwCRPS employs the weight function $w(\alpha) = \one_{\{\alpha > 0.99\}}$, focusing on the upper tail at quantile level $\tau = 0.99$. Columns correspond to weather variables and rows correspond to lead times. Skill is measured relative to the $\overline{\operatorname{qwPCRPS}}_{0,\tau}$ baseline. Brighter colours indicate higher potential skill relative to this baseline, while darker colours indicate lower potential skill.}
	\label{fig:maps_fuxi_qw_pcrpss_99_vs_era5}
\end{figure}

The relative improvement in $\overline{\operatorname{qwPCRPS}}_\tau$ of FuXi upon the unconditional climatology is shown at each grid point in Figure \ref{fig:maps_fuxi_qw_pcrpss_99_vs_era5} when interest is on the upper tail of the forecast distributions. The conclusions are qualitatively similar to those for the $\operatorname{twPCRPS}$ in the main text. Improvements are lowest over the tropics and largest over the extratropics for all variables, with the skill of the FuXi forecasts decreasing as lead time increases. However, since the 99th percentile of the forecast distribution typically corresponds to a less extreme event than the 99th percentile of the past observations, FuXi retains skill even at lead times of 10 days when assessed using the $\operatorname{qwPCRPS}$, which is not the case when assessed using the $\operatorname{twPCRPS}$.

\begin{figure}[t!]
	\centering
	\includegraphics[width=\linewidth]{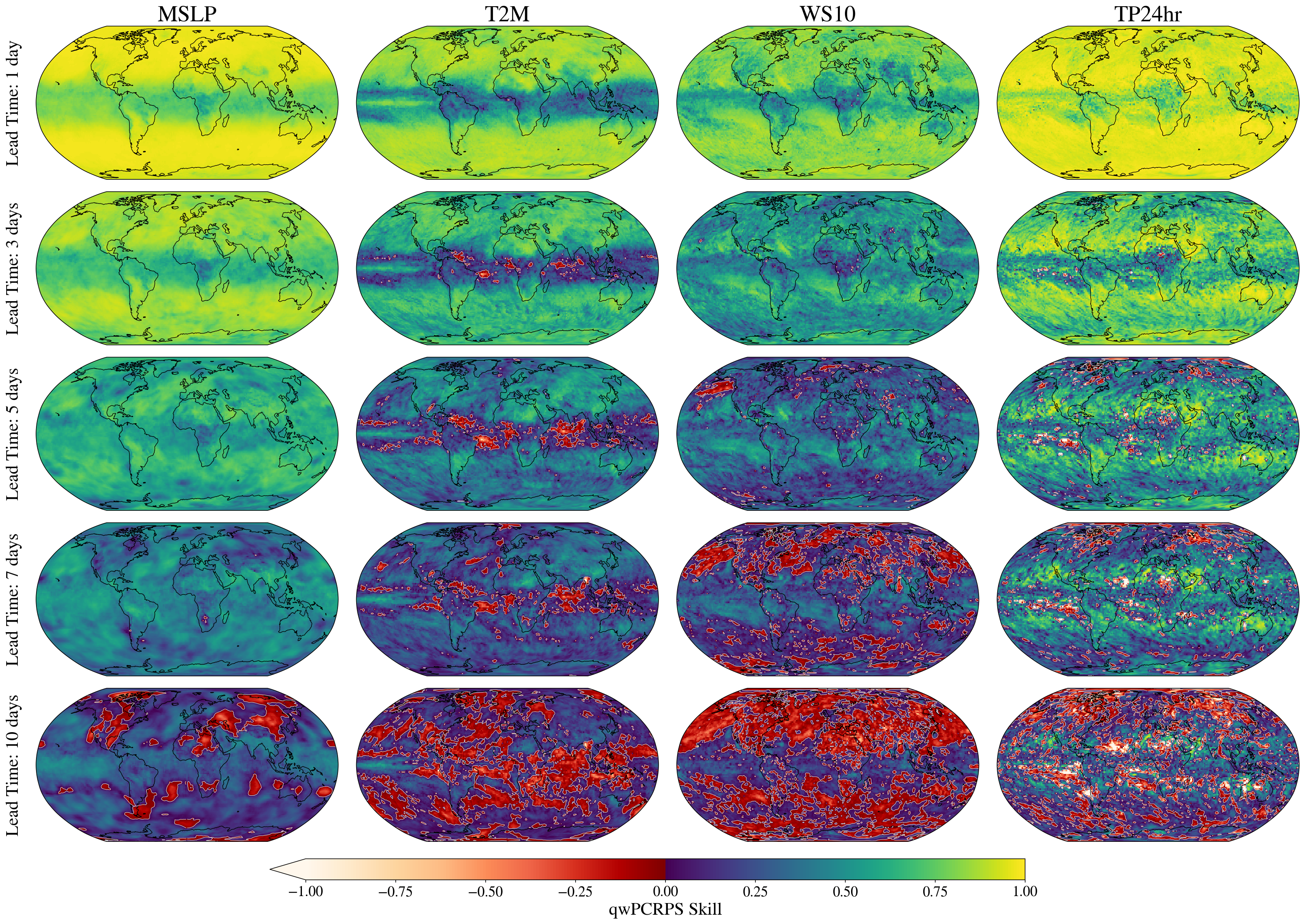}
	\caption{$\operatorname{qwPCRPS}_\tau$ skill of FuXi relative to the seasonally varying ERA5 climatology forecast from WeatherBench 2. Results are shown at each grid point using ERA5 reanalyses as verification data. The qwCRPS employs the weight function $w(\alpha) = \one_{\{\alpha > 0.99\}}$, focusing on the upper tail at quantile level $\tau = 0.99$. Columns correspond to weather variables and rows correspond to lead times. Positive values indicate improvement over the climatological baseline, while negative values indicate worse performance.}
	\label{fig:maps_fuxi_qw_pcrps_skill_99_vs_era5}
\end{figure}

However, when FuXi is compared with a more skilful baseline, such as a seasonally varying climatology forecast, then its skill decreases at a much faster rate. Figure \ref{fig:maps_fuxi_qw_pcrps_skill_99_vs_era5} displays the skill of FuXi relative to the seasonally varying ERA5 climatology forecast from WeatherBench 2, with the forecasts again assessed using the $\overline{\operatorname{qwPCRPS}}_\tau$ with $\tau = 0.99$. A similar pattern is observed to Figure \ref{fig:maps_fuxi_qw_pcrpss_99_vs_era5} at a lead time of one day, but the seasonal climatology outperforms FuXi after just three days when predicting temperature at many grid points in the tropics. At a lead time of 10 days, the skill is weak or negative across large parts of the globe for all variables, indicating that FuXi often does not improve upon the seasonal climatology at this lead time. This is particularly pronounced for wind speed, where FuXi performs worse than the seasonal baseline at the majority of grid points, especially in the extratropics.

\begin{figure}[t!]
	\centering
	\includegraphics[width=\linewidth]{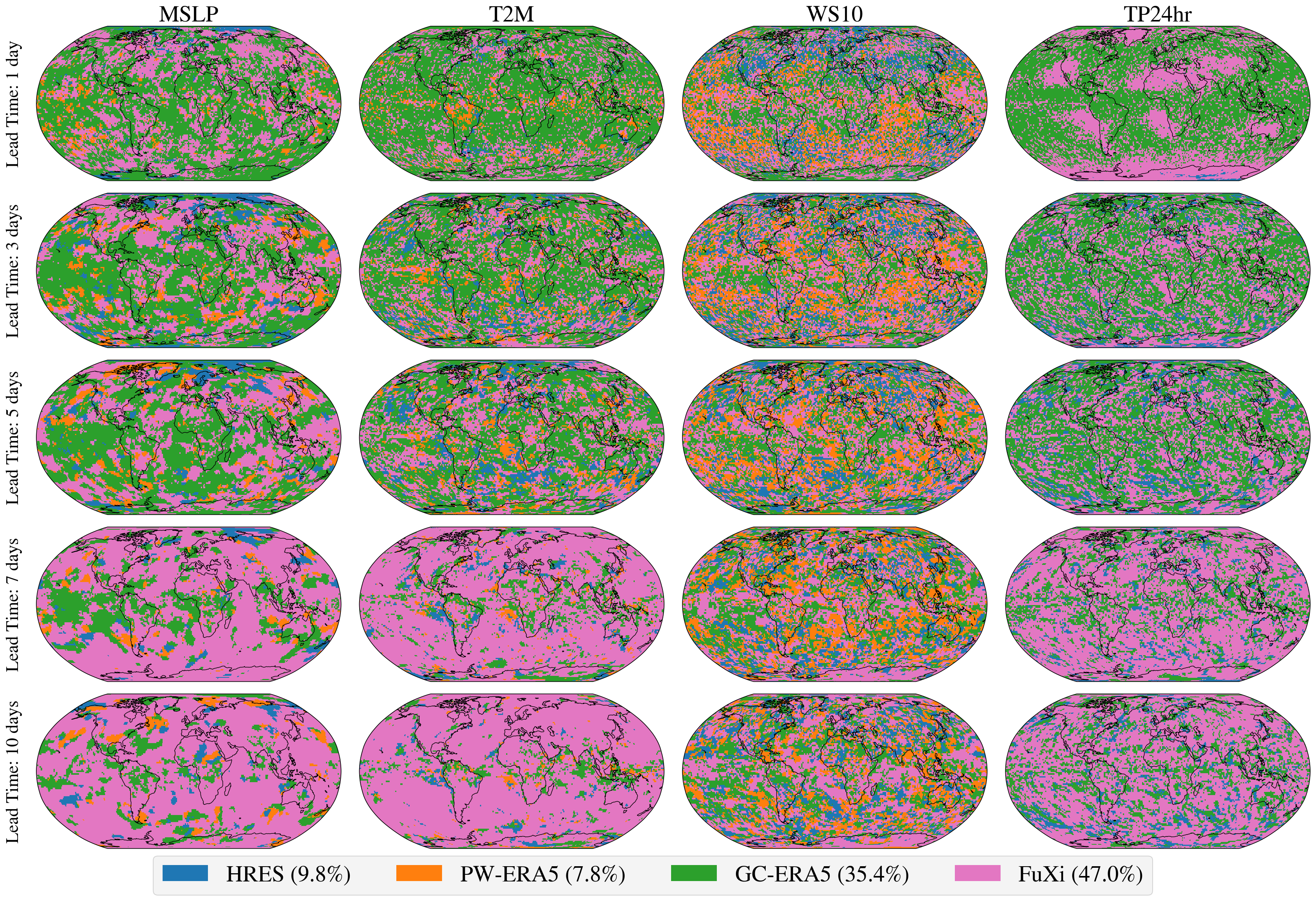}
	\caption{Best-performing model according to $\overline{\operatorname{qwPCRPS}}_\tau$ with weight function $w(\alpha) = \one_{\{\alpha > 0.99\}}$, focusing on the upper tail at quantile level $\tau = 0.99$. Results are shown at each grid point using ERA5 reanalyses as verification data. The colour at each grid point indicates the model with the lowest $\overline{\operatorname{qwPCRPS}}_\tau$. Columns correspond to weather variables and rows correspond to lead times. PW-ERA5 is not available for TP24hr and is therefore not present in the final column. The percentages in the legend correspond to the total proportion of cases that each model performs best, aggregated across all grid points, variables, and lead times.}
	\label{fig:maps_best_model_qw_pcrps_99_vs_era5}
\end{figure}

\begin{figure}[t!]
	\centering
	\includegraphics[width=\linewidth]{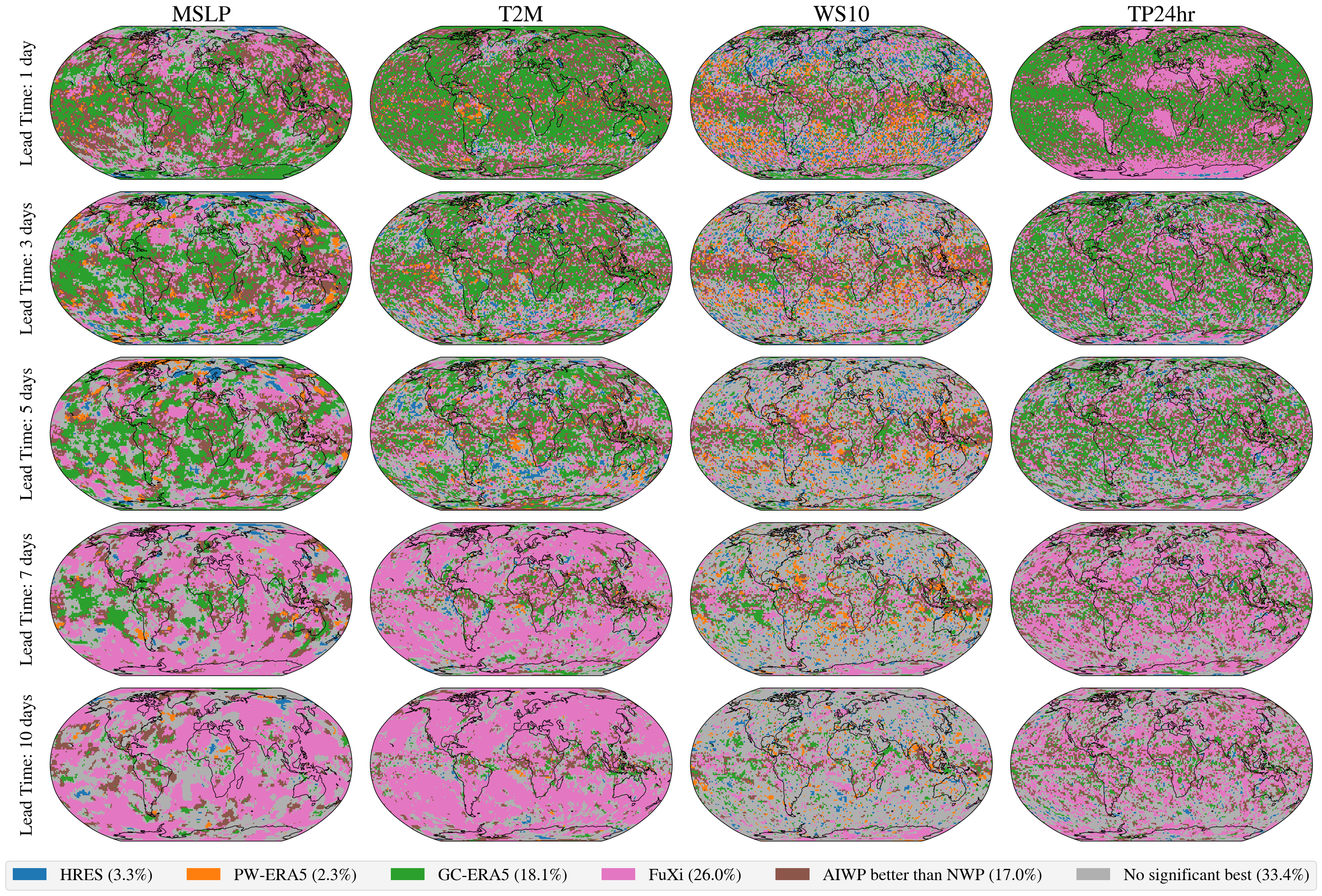}
	\caption{Significance of the best-performing model according to the $\overline{\operatorname{qwPCRPS}}_\tau$ with weight function $w(\alpha) = \one_{\{\alpha > 0.99\}}$, focusing on the upper tail at quantile level $\tau = 0.99$. Results are shown at each grid point using ERA5 reanalyses as verification data. The colour at each grid point indicates whether a model receives a $\overline{\operatorname{qwPCRPS}}_\tau$ that is significantly lower than that of all other available models at a 5\% significance level. Grid points for which all available AIWP models significantly outperform HRES, but no individual AIWP model significantly outperforms the others, are indicated in brown. Grey grid points indicate locations where no significantly best model is identified. Columns correspond to weather variables and rows correspond to lead times. PW-ERA5 is not available for TP24hr and is therefore not present in the final column.}
	\label{fig:maps_sig_best_model_ai_qw_pcrps_monthly_max_vs_era5}
\end{figure}

At short lead times, GraphCast appears to provide the most accurate forecasts in terms of $\operatorname{qwPCRPS}$ at the majority of grid points, while FuXi is clearly preferable at longer lead times (Figure \ref{fig:maps_best_model_qw_pcrps_99_vs_era5}). This is the case for MSLP, T2M, and TP24hr, and is most pronounced when forecasting temperature. A similar result was found for the $\operatorname{twPCRPS}$ in the main text. However, in contrast to the results for the $\operatorname{twPCRPS}$, these differences in predictive performance with respect to the $\operatorname{qwPCRPS}$ are generally often statistically significant, as indicated by Figure \ref{fig:maps_sig_best_model_ai_qw_pcrps_monthly_max_vs_era5}; these hypothesis tests are performed using the block permutation test described in Appendix~B of the main manuscript. As mentioned above, the reason for this is that we are essentially focusing on less extreme events when employing the quantile-weighted CRPS, which leads to less uncertainty in the score estimates. In contrast to the other variables, no model is clearly preferred when forecasting wind speed at any lead time, with all four models resulting in the best-performing forecasts at several grid points. Perhaps unsurprisingly, Figure \ref{fig:maps_sig_best_model_ai_qw_pcrps_monthly_max_vs_era5} suggests that the differences between models in this case are often not statistically significant, particularly at longer lead times.

\end{document}